\documentclass[preprint,12pt]{elsarticle}

\biboptions{sort&compress}

\usepackage[T1]{fontenc}
\usepackage[utf8]{inputenc}
\usepackage{lmodern}
\usepackage{amsmath,amssymb}
\usepackage{graphicx}
\usepackage{longtable,booktabs,array}
\usepackage{calc}
\usepackage{etoolbox}
\usepackage{url,xurl}
\usepackage{microtype}
\usepackage{textcomp}
\usepackage{float}

\providecommand{\real}[1]{#1}

\makeatletter
\patchcmd\longtable{\par}{\if@noskipsec\mbox{}\fi\par}{}{}
\makeatother
\AtBeginEnvironment{longtable}{\small\sloppy}
\journal{Progress in Particle and Nuclear Physics}

\begin{document}
\begin{frontmatter}
\title{Indirect Dark Matter Detection in the Multi-Messenger Era: Experiments, Systematics, Anomalies,\\ and Prospects for Discovery}
\author[queens,zu]{Muad Ghaith\corref{cor1}}
\ead{muad.ghaith@queensu.ca}
\cortext[cor1]{Corresponding author}
\author[ukf]{Khaled Obaideen}
\address[queens]{Department of Physics, Queen's University, Kingston, ON K7L 3N6, Canada}
\address[zu]{College of Natural and Health Sciences, Zayed University, Dubai 19282, U.A.E}
\address[ukf]{College of Computing and Intelligent Systems, University of Khorfakkan, Khorfakkan, U.A.E}

\begin{abstract}
Indirect searches for dark matter (DM) span gamma rays, charged cosmic rays, neutrinos, X-rays, radio emission, the cosmic microwave background (CMB), and the 21~cm signal, yet the evidential strength of reported constraints and anomalies varies substantially across messengers. This critical review synthesizes peer-reviewed literature published from January 2000 through October 2026 using a transparent, structured search strategy.

We assess observations within a common framework that links particle interactions and halo source terms to propagation or cosmological energy deposition, instrumental response, background modeling, and statistical inference. Fermi-LAT observations of dwarf spheroidal galaxies provide one of the most robust GeV-scale benchmarks for DM annihilation, whereas Galactic-center searches offer higher expected signal intensities but are more sensitive to foreground modeling and halo profiles. Charged-cosmic-ray measurements achieve high precision but remain limited by propagation and source uncertainties, while neutrino observations provide complementary sensitivity through solar capture, directional signatures, flavor information, and searches for high-mass annihilation or decay. Several widely discussed signals---including the 3.5~keV line, 511~keV emission, the Galactic-center GeV excess, charged-particle features, cluster radio excesses, and candidate gamma-ray lines---remain inconclusive because of limited independent confirmation, uncertain target scaling, spectral degeneracies, or background-modeling limitations.

Cosmological observations provide complementary constraints from CMB energy-injection limits and 21~cm measurements of the intergalactic medium. Artificial intelligence increasingly supports reconstruction and inference but requires rigorous calibration and domain-shift validation. The strongest case for DM discovery would require statistically consistent evidence across independent targets, instruments, or messengers, under common particle-physics parameters, and with well-characterized uncertainties.
\end{abstract}
\end{frontmatter}

\section{Introduction: From Sensitivity to Credibility}\label{introduction-from-sensitivity-to-credibility}

Dark matter is established through gravitational evidence, yet its microscopic nature remains unknown. Indirect searches seek non-gravitational signatures from annihilation, decay, conversion, or Hawking emission. Their strength lies in the enormous astrophysical targets and baselines unavailable in laboratories; their challenge is that observable signals depend not only on particle physics but also on DM distributions, radiation and magnetic fields, particle transport, cosmological energy deposition, detector response, and background modeling. Consequently, translating experimental sensitivity into discovery credibility remains a central challenge \cite{ref1,ref2,ref3,ref4,ref5}.

This review addresses this challenge across gamma rays, charged cosmic rays, neutrinos, X-rays, radio emission, the cosmic microwave background (CMB), and the 21-cm signal within a unified inference framework. Robust indirect evidence requires candidate hypotheses to remain consistent across independent targets, energy regimes, and nuisance models. Individual limits may be powerful but model-dependent, while excesses can persist without being DM-specific. Likewise, combined analyses may overstate significance if they neglect shared astrophysical priors or correlated uncertainties. A credible discovery therefore requires moving from an unexplained residual to a reproducible, independently testable prediction.

\begin{figure}[htbp]
\centering
\includegraphics[width=0.95\linewidth]{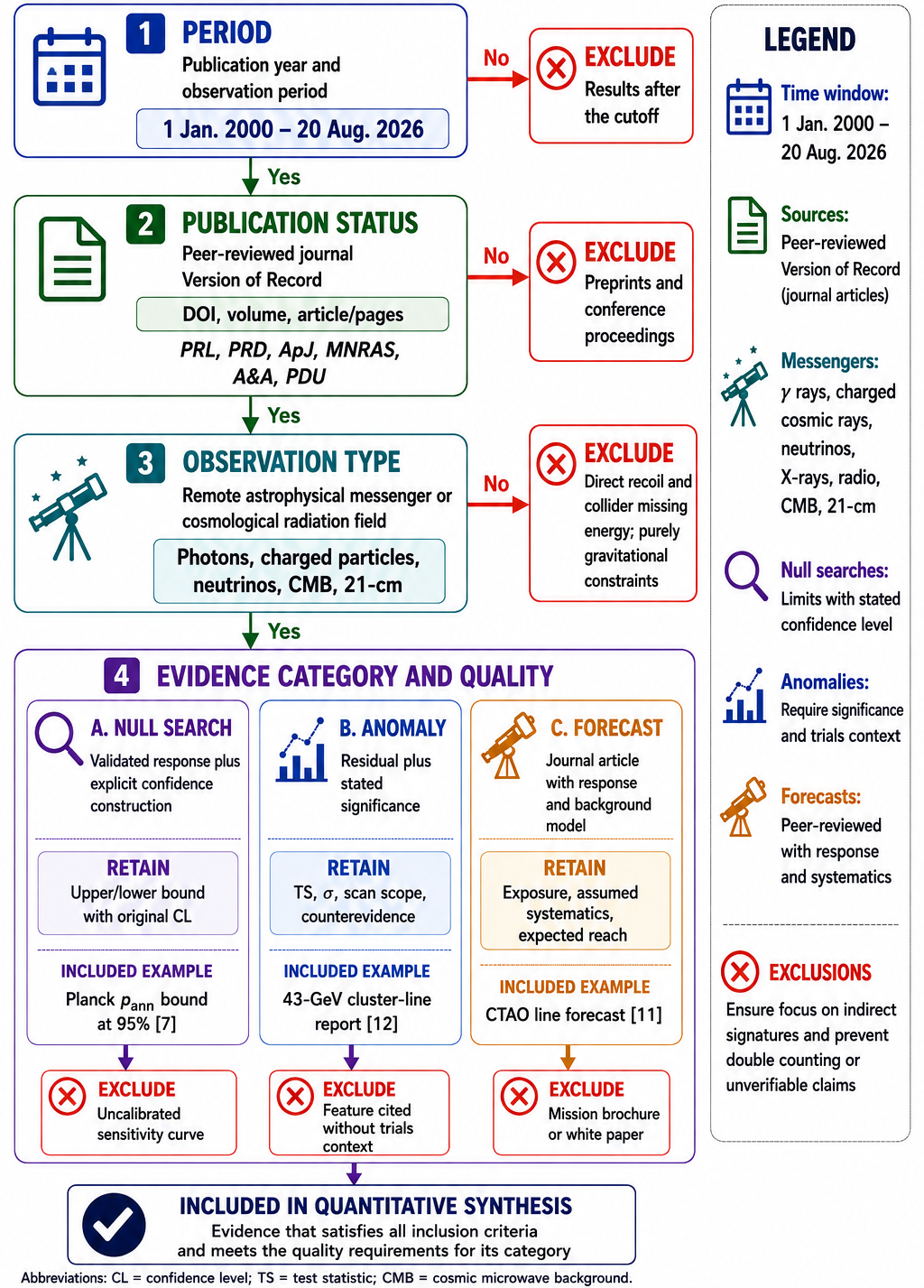}
\caption{Operational framework for including and classifying evidence as null searches, anomalies, or forecasts. We excluded direct detection, collider searches, gravitational-only constraints, and non-peer-reviewed or inadequately characterized results. Sources: \cite{ref6,ref7,ref8,ref9,ref10,ref11}.}
\label{fig:operational-framework}
\end{figure}

Fig.~\ref{fig:operational-framework} shows the scope of this review. We include astrophysical observations with a quantifiable connection to DM processes, while excluding direct-detection experiments, collider searches, and purely gravitational constraints from the quantitative synthesis. This delineation keeps the comparison focused on indirect signatures carried by photons, charged particles, neutrinos, and cosmological radiation fields.

These criteria distinguish breadth of coverage from strength of evidence. The scope is process-based: the review includes observations in which dark-sector processes produce or modify remotely observed astrophysical messengers or cosmological radiation fields. We therefore include solar-neutrino searches and CMB and 21-cm constraints, while excluding terrestrial boosted-DM recoil measurements from the quantitative synthesis. We use direct-detection and collider results only as external context.

The framework distinguishes departures from a background-only hypothesis from evidence favoring a DM interpretation. Null searches can yield strong constraints, whereas any residual must withstand tests against alternative background models. Spectral, morphological, cross-target, multi-messenger, and
independent instrumental consistency provide increasingly strong validation. We do not assign numerical evidence grades because discovery credibility depends on several correlated factors.

Table~\ref{tab:dm_candidates} provides an overview of candidates that extends beyond conventional weakly interacting massive particle (WIMP) annihilation to sterile neutrinos, axion-like particles (ALPs), dark photons, primordial black holes (PBHs), secluded sectors, and superheavy relics. Despite
their distinct signatures and systematic uncertainties, these candidates can be compared using a common inference chain comprising source normalization, propagation or conversion, detector response, background modeling, and statistical calibration.


\setlength{\LTcapwidth}{\columnwidth}
\setlength{\tabcolsep}{2pt}
\renewcommand{\arraystretch}{1.15}
\setlength{\emergencystretch}{2em}

\footnotesize

\begin{longtable}[]{@{}
  >{\raggedright\arraybackslash}p{(\columnwidth - 10\tabcolsep) * \real{0.14}}
  >{\raggedright\arraybackslash}p{(\columnwidth - 10\tabcolsep) * \real{0.16}}
  >{\raggedright\arraybackslash}p{(\columnwidth - 10\tabcolsep) * \real{0.18}}
  >{\raggedright\arraybackslash}p{(\columnwidth - 10\tabcolsep) * \real{0.17}}
  >{\raggedright\arraybackslash}p{(\columnwidth - 10\tabcolsep) * \real{0.13}}
  >{\raggedright\arraybackslash}p{(\columnwidth - 10\tabcolsep) * \real{0.22}}
@{}}

\caption{Representative DM candidates and indirect-detection processes, showing characteristic source scaling, primary signatures, parameter ranges, observable energy domains, and dominant uncertainties \cite{ref12,ref13,ref14,ref15,ref16,ref17,ref18,ref19,ref20,ref21,ref22,ref23,ref24,ref25,ref26,ref27,ref28,ref29,ref30,ref31,ref32,ref33,ref34,ref35,ref36,ref37}.}
\label{tab:dm_candidates} \\

\toprule
\textbf{Candidate / process} &
\textbf{Source scaling} &
\textbf{Indirect signature} &
\textbf{Candidate parameter} &
\textbf{Energy domain} &
\textbf{Dominant uncertainty} \\
\midrule
\endfirsthead

\toprule
\textbf{Candidate / process} &
\textbf{Source scaling} &
\textbf{Indirect signature} &
\textbf{Candidate parameter} &
\textbf{Energy domain} &
\textbf{Dominant uncertainty} \\
\midrule
\endhead

\midrule
\multicolumn{6}{r}{\footnotesize Continued on next page} \\
\endfoot

\bottomrule
\endlastfoot


WIMP annihilation &
\(\rho^{2}\langle \sigma v\rangle/m_{\chi}^{2}\) &
\(\gamma\)-ray continuum/lines; antimatter; neutrinos &
\(m_{\chi}\): GeV--PeV &
MeV--PeV &
\(J\)-factor; annihilation channels and branching fractions \\

DM decay &
\(\rho/(m_{\chi}\tau_{\chi})\) &
Diffuse \(\gamma\) rays; neutrinos; X-ray lines &
\(m_{\chi},\tau_{\chi}\): model dependent &
keV--ZeV &
\(D\)-factor; Galactic/extragalactic contributions \\

Sterile-neutrino decay &
\(\rho\,\sin^{2}(2\theta)m_{s}^{5}\) &
Monochromatic X-ray, \(E_{\gamma}\simeq m_{s}/2\) &
\(m_{s}\sim2\)--20 keV; mixing angle \(\theta\) &
\(\sim1\)--10~keV &
Atomic-line backgrounds; DM column density \\

ALP conversion &
Magnetic-domain conversion probability&
Spectral irregularities; photon/radio features&
\(m_{a}\sim\) neV--keV; coupling \(g_{a\gamma}\) &
Source dependent&
Magnetic-field strength, geometry, and coherence \\

Dark-photon conversion&
Plasma resonance \(\times\) kinetic mixing&
Radio emission; CMB distortions&
Ultralight \(m_{A'}\); kinetic mixing \(\varepsilon\) &
Radio--microwave &
Plasma and electron-density structure \\

PBH evaporation &
Mass-function-weighted Hawking emission &
\(\gamma\) rays; \(e^{\pm}\); neutrinos &
\(M_{\mathrm{PBH}}\sim10^{15}\)--\(10^{18}\,\mathrm{g}\) &
MeV--GeV\(^*\)&
PBH mass function; Galactic distribution \\

Secluded annihilation&
Capture/halo rate \(\times\) mediator survival&
Solar \(\gamma\) rays; neutrinos&
\(m_{\chi}\), mediator mass/lifetime &
GeV--PeV&
Mediator lifetime, propagation, and absorption \\

Superheavy DM decay&
\(\rho/(m_{\chi}\tau_{\chi})\) + particle shower&
UHE \(\gamma\) rays; neutrinos&
\(m_{\chi}\): superheavy regime; \(\tau_{\chi}\)&
TeV--ZeV&
Electroweak cascades; propagation and attenuation \\

\midrule
\multicolumn{6}{@{}p{\columnwidth}@{}}{%
\footnotesize
\textit{*The observable spectrum from PBH evaporation depends strongly on PBH mass and the treatment of primary and secondary Hawking-emission products.}
} \\

\end{longtable}

\normalsize

Candidate diversity is important because each DM scenario has distinct falsifiable predictions and systematic uncertainties. WIMP constraints depend on annihilation channels, sterile-neutrino lines on atomic backgrounds, and ALP or dark-photon conversion on independently measurable magnetic and plasma environments. Primordial black hole (PBH) searches depend strongly on the assumed mass function, while secluded sectors introduce mediator propagation and superheavy DM decay involves electroweak cascades and anisotropy. Consequently, discovery credibility cannot be defined by statistical significance alone but must account for candidate-specific predictions and background uncertainties.

Messenger complementarity similarly depends on propagation and nuisance orthogonality. Gamma rays and neutrinos largely preserve directional information, though attenuation and absorption can matter. Charged cosmic rays are strongly transport-dependent, whereas radio and X-ray secondaries depend on magnetic and radiation fields. Cosmological probes sacrifice spatial localization for sensitivity across cosmic time. Complementary channels are therefore most informative when they share particle-physics predictions but have largely independent systematic uncertainties.

\begin{figure}[htbp]
\centering
\includegraphics[width=1\linewidth]{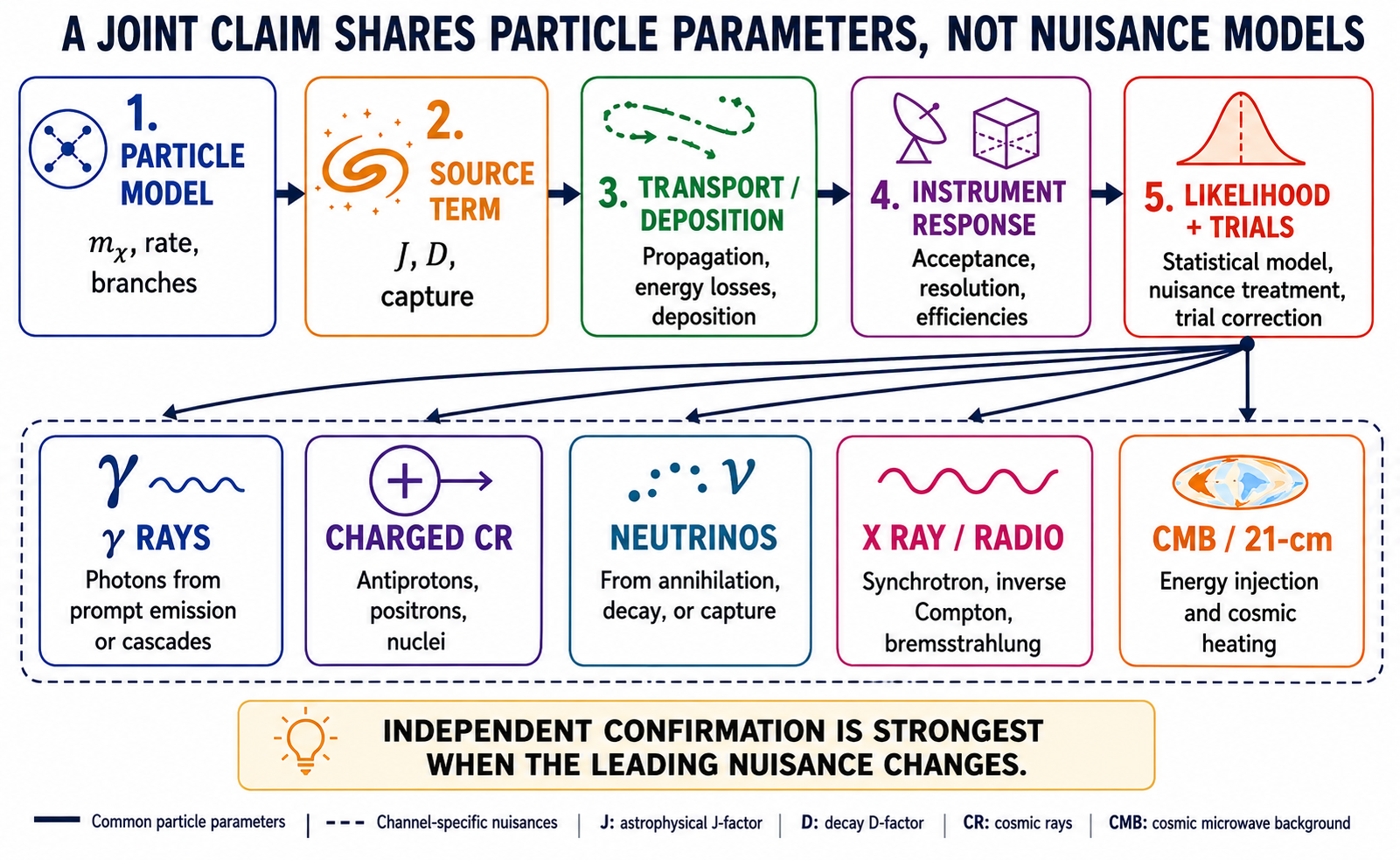}
\caption{Source-to-inference architecture for indirect detection. The diagram identifies the shared particle parameters and the messenger-specific transport, response, background, and statistical layers that must remain explicit in a joint analysis. Sources: \cite{ref7,ref12,ref14}.}
\label{fig:source-inference}
\end{figure}

Fig.~\ref{fig:source-inference} summarizes this common inference architecture. Each messenger links a DM source to propagation or energy deposition, detector response, background modeling, and statistical inference. Thus, experimental comparisons should consider not only the strength of an upper limit but also which elements of this inference chain are empirically constrained and which remain model-dependent assumptions.

The energy domains covered by indirect-detection messengers do not share a common sensitivity scale: performance depends on signal spectra, detector thresholds, event selection, and energy resolution. Coverage remains comparatively limited in the sub-MeV-to-GeV regime, affecting light-DM and primordial black hole (PBH) searches, while the broad neutrino range does not imply uniform sensitivity because effective areas, fluxes, and backgrounds vary strongly with energy. Candidate models provide a complementary mapping: weakly interacting massive particles (WIMPs) can produce signals across multiple messengers, sterile-neutrino decay predicts a characteristic line constrained by DM column density and atomic backgrounds, axion-like particle (ALP) and dark-photon signals depend on magnetic or plasma conversion, and PBH signatures depend strongly on the mass function. Consequently, independent targets and messengers provide stronger tests when a model consistently predicts both detections and non-detections.

\section{Review Methodology and Evidence Assessment}\label{review-methodology-and-evidence-grading}

This critical narrative review used a transparent, structured search covering January 2000 to August 2026. Searches combined DM terminology with major messengers, candidate models, targets, anomalies, machine learning, and systematics across NASA ADS, INSPIRE, Web of Science, Scopus, IEEE Xplore-compatible searches, and publisher/Crossref records. Citation tracing supplemented database searches. Given the heterogeneity of messengers, targets, observables, and likelihoods, we did not perform a pooled meta-analysis.

We included peer-reviewed journal articles reporting astrophysical observations, observational analyses, or response-calibrated forecasts relevant to near-term experiments. We excluded preprints, conference proceedings, white papers, mission materials, direct-detection and collider searches, and purely gravitational constraints. When a journal article superseded a preprint, we retained only the Version of Record.

For each study, we extracted the messenger, target, candidate/process, exposure, energy or mass range, astrophysical assumptions, confidence construction, quantitative result, uncertainty, and dominant systematic.
We preserved reported confidence levels and identified values extracted from figures as approximate. We distinguished directly reported results from quantities reinterpreted in this review.

We classified evidence as calibrated null searches, residuals, spectral features, morphological matches, cross-target consistency, multi-messenger predictions, or replicated signals. These categories indicate evidential meaning rather than numerical quality. A residual shows the baseline model's inadequacy but does not establish a DM origin, whereas well-calibrated null results can provide robust constraints and reusable likelihood information.

Interpretation followed three principles: assumptions are integral to reported results; local significance does not alone establish discovery credibility; and cross-target or multi-messenger tests should be predictive rather than independently retuned. We applied these
principles consistently to both reported anomalies and exclusion limits.

\section{Observable Framework: Source Terms, Halos, and Targets}\label{observable-framework-source-terms-halos-and-targets}

\subsection{Conceptual Flux Equations}\label{conceptual-flux-equations}

For self-conjugate DM of mass \(m_{\chi}\), the prompt differential flux from annihilation is
\begin{equation}
\frac{d\Phi_{\mathrm{ann}}}{dE}
=
\frac{\left\langle \sigma v \right\rangle}
{8\pi m_{\chi}^{2}}
\sum_{f} B_{f}\frac{dN_{f}}{dE}\,J,
\end{equation}

where the astrophysical \(J\)-factor is

\begin{equation}
J
=
\int_{\mathrm{ROI}} d\Omega
\int_{\mathrm{los}} \rho^{2}(r)\,dl.
\end{equation}

\noindent while for decay,
\begin{equation}
\frac{d\Phi_{\mathrm{dec}}}{dE}
=
\frac{1}{4\pi m_{\chi}\tau_{\chi}}
\sum_{f} B_{f}\frac{dN_{f}}{dE}\,D,
\end{equation}

where the astrophysical \(D\)-factor is

\begin{equation}
D
=
\int_{\mathrm{ROI}} d\Omega
\int_{\mathrm{los}} \rho(r)\,dl.
\end{equation}

These expressions separate particle-physics parameters from the astrophysical \(J\)- and \(D\)-factors, although realistic predictions may additionally require propagation, secondary emission, absorption, detector response, oscillations, or cosmological evolution. Cross-study comparisons therefore require consistent spectra, regions of interest (ROIs), density profiles, and statistical conventions \cite{ref7,ref12}.

\begin{figure}[htbp]
\centering
\includegraphics[width=0.95\linewidth]{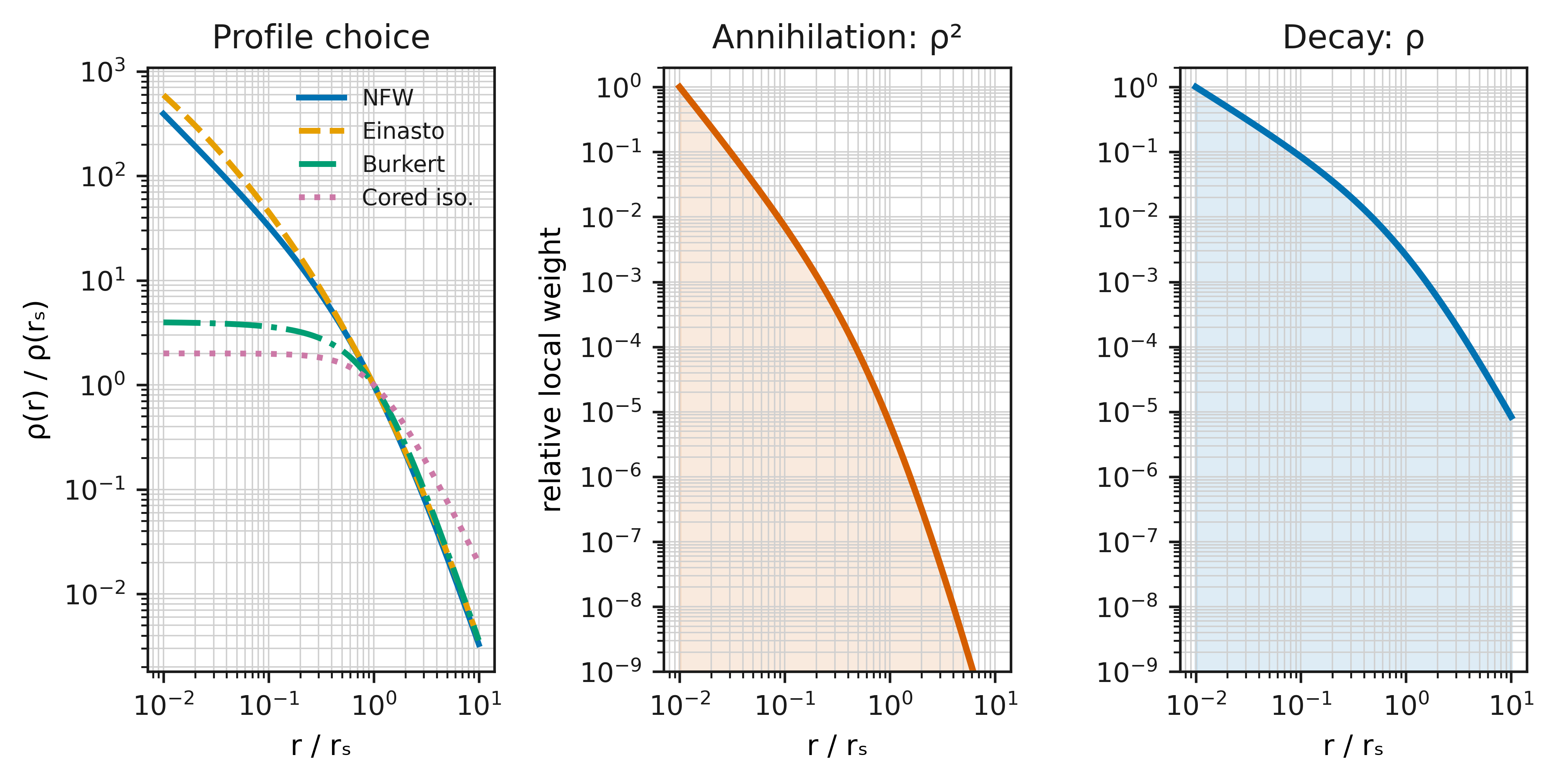}
\caption{Density profile and target leverage. The left panels compare normalized NFW, Einasto, Burkert, and cored profiles; the right panel shows how $\rho$ and $\rho^2$ weighting change the concentration of decay and annihilation signals. Curves are analytic illustrations, not fitted Milky Way profiles. Sources: \cite{ref40,ref41,ref42,ref43}.}
\label{fig:density-profile}
\end{figure}

Fig.~\ref{fig:density-profile} illustrates the contrasting density dependence.
Annihilation scales as \(\rho^{2}\), enhancing emission from dense central regions but increasing sensitivity to halo-profile and background uncertainties. Decay scales as \(\rho\), producing a broader morphology and generally weaker dependence on the central density profile. Thus, higher nominal annihilation flux does not necessarily imply greater robustness.

Unresolved subhalos can enhance cluster annihilation emission, but the inferred boost depends strongly on extrapolations of the subhalo population and mass--concentration relation below resolved simulation scales. Target rankings therefore vary with astrophysical priors even for unchanged gamma-ray data. Velocity-dependent annihilation (including $p$-wave or Sommerfeld-enhanced scenarios) further limits the applicability of a constant $\langle\sigma v\rangle$ across environments. Before structure formation, smooth-density energy injection scales approximately as $(1+z)^6$ for annihilation and $(1+z)^3$ for decay; structure formation modifies these predictions at late times \cite{ref6,ref13,ref40,ref41,ref42,ref43,ref45,ref46}.

The different density scalings also affect control-region selection, angular morphology, and substructure contributions. Near the Galactic center, the annihilation signal can extend into background regions, requiring the assumed morphology to be incorporated into both signal and control regions. Decay produces a broader halo component that can be partially degenerate with isotropic or large-scale diffuse emission. In clusters, substructure boosts are spatially extended and can modify the optimal region of interest, so treating the boost as a single multiplicative factor neglects relevant spatial information. Similar distinctions arise cosmologically, where annihilation and decay probe different redshift and density regimes, with clumping becoming important only after structure formation.

Accordingly, annihilation and decay should be compared at the count-map or likelihood level where possible, rather than through aperture-integrated flux limits alone. Treat astrophysical uncertainties according to their physical origin. Milky Way halo uncertainties depend on the baryonic potential, circular velocity, bar geometry, and equilibrium assumptions; dwarf-galaxy factors depend on stellar membership and dynamical modeling; and cluster boosts depend on unresolved subhalo populations. These distinct uncertainties should not
be represented by a single generic nuisance prescription. Robust analyses should therefore report Galactic-center sensitivity versus core size, propagate non-Gaussian dwarf \(J\)- or \(D\)-factor posteriors, and separate smooth and substructure components in cluster analyses. More generally, astrophysical inference should remain traceable to the observations constraining each halo parameter and its correlation with
the adopted region of interest. Otherwise, statistically precise particle-physics limits may remain astrophysically overconstrained.

\subsection{Halo Profiles and Dwarf Factors}\label{halo-profiles-and-dwarf-factors}

Halo modeling and target selection are closely linked because the preferred target depends on both the assumed DM distribution and the dominant astrophysical uncertainties. NFW and Einasto profiles serve as useful baselines but should not be treated as empirical descriptions at arbitrarily small radii. Baryonic contraction, stellar feedback, bars, and dynamical history can steepen or flatten the inner Milky Way profile. Their convergence at large radii makes broad decay searches comparatively robust, whereas their divergence toward the center produces substantial variation in Galactic-center annihilation limits and can alter the optimal region of interest.

Dwarf spheroidal galaxies offer a cleaner alternative because stellar kinematics constrain their DM distributions. However, their \(J\)-factor posteriors remain sensitive to limited membership samples, unresolved binaries, velocity anisotropy, triaxiality, and tidal effects \cite{ref40,ref41,ref42}. These uncertainties should be propagated hierarchically, particularly in combined analyses, to prevent a poorly
constrained ultrafaint system from dominating the likelihood.

No single target simultaneously maximizes signal strength, background rejection, spatial information, and astrophysical robustness. The Galactic center generally offers a strong annihilation signal but has substantial foreground and halo-profile uncertainties, whereas dwarf galaxies provide cleaner backgrounds at lower expected flux. Galaxy clusters probe large DM masses but depend on substructure boosts and
cosmic-ray emission. Extragalactic searches sample large cosmological volumes but introduce uncertainties from unresolved source populations and structure formation. Solar searches offer a localized directional signal but depend on capture and scattering physics, while nearby galaxies provide independent environments suitable for morphological
tests.

Table~\ref{tab:dm_targets_comparison} illustrates why target selection is a multidimensional optimization rather than a fixed ranking. For example, if a cuspy profile favors a Galactic-center signal, test it against dwarf galaxies and other independent probes if cored profiles are allowed. More generally, complementary targets should
constrain different astrophysical nuisance parameters or distinct consequences of the same particle model. Compare Galactic-center WIMP interpretations with dwarf and CMB constraints; compare sterile-neutrino line candidates with independent low-background targets or higher-resolution spectroscopy; compare ALP scenarios with sources that sample different magnetic environments; and compare secluded solar models with joint gamma-ray and neutrino observations.

Observing strategies should maximize information gain across plausible nuisance scenarios rather than optimize a single benchmark upper limit. Balanced target portfolios, improved stellar spectroscopy of dwarf galaxies, preservation of control regions, and cross-target consistency tests provide a stronger basis for discovery, exclusion, background
characterization, and independent validation. The resulting trade-offs among target leverage, backgrounds, and independent validation tests are summarized in Table~\ref{tab:dm_targets_comparison}.

\setlength{\LTcapwidth}{\columnwidth}
\setlength{\tabcolsep}{2pt}
\renewcommand{\arraystretch}{1.15}
\setlength{\emergencystretch}{2em}

\footnotesize

\begin{longtable}[]{@{}
  >{\raggedright\arraybackslash}p{(\columnwidth - 8\tabcolsep) * \real{0.13}}
  >{\raggedright\arraybackslash}p{(\columnwidth - 8\tabcolsep) * \real{0.24}}
  >{\raggedright\arraybackslash}p{(\columnwidth - 8\tabcolsep) * \real{0.16}}
  >{\raggedright\arraybackslash}p{(\columnwidth - 8\tabcolsep) * \real{0.25}}
  >{\raggedright\arraybackslash}p{(\columnwidth - 8\tabcolsep) * \real{0.22}}
@{}}

\caption{Comparative strengths, dominant systematics, and complementary tests for major astrophysical targets used in indirect DM searches. Target selection reflects a trade-off among expected signal strength, background complexity, spatial information, and robustness of the underlying DM distribution. Sources: \cite{ref8,ref45,ref48,ref49,ref50}.}
\label{tab:dm_targets_comparison} \\

\toprule
\textbf{Target} &
\textbf{Observational / statistical leverage} &
\textbf{Sensitive messenger(s)} &
\textbf{Dominant background / systematic} &
\textbf{Complementary test} \\
\midrule
\endfirsthead

\toprule
\textbf{Target} &
\textbf{Observational / statistical leverage} &
\textbf{Sensitive messenger(s)} &
\textbf{Dominant background / systematic} &
\textbf{Complementary test} \\
\midrule
\endhead

\midrule
\multicolumn{5}{r}{\footnotesize Continued on next page} \\
\endfoot

\bottomrule
\endlastfoot


Galactic center &
Largest nominal annihilation \(J\)-factor; analyses commonly focus on
the inner \(\sim1^{\circ}\)--\(20^{\circ}\), depending on instrument
and background model. &
Gamma rays; high-energy neutrinos &
Diffuse Galactic emission, unresolved sources, and uncertainty in the
inner DM profile. &
Dwarf spheroidals, Galactic halo, CMB constraints, and independent
instruments. \\

Dwarf spheroidal galaxies &
Low astrophysical backgrounds and stellar-kinematic constraints on
individual \(J\)- and \(D\)-factors; 15 dwarfs were included in the
2015 Fermi-LAT joint analysis. &
GeV gamma rays; TeV gamma rays and neutrinos for complementary mass
ranges &
Stellar membership, velocity-anisotropy assumptions, limited stellar
samples, and non-Gaussian \(J\)-factor posteriors. &
IACT stacking, neutrino stacking, and comparison with Galactic-center
constraints. \\

Galaxy clusters &
Large DM mass and potential enhancement from unresolved substructure;
16 clusters were considered in a 2025 Fermi-LAT analysis. &
Gamma rays, radio emission, neutrinos, and decay-line searches &
AGN emission, cosmic rays, intracluster emission, halo modeling, and
uncertain substructure boost. &
Dwarf galaxies, Galactic halo/center analyses, and multiwavelength
observations. \\

Galactic halo &
Large solid angle and substantial integrated DM column; particularly
favorable for decay because of the broad \(D\)-factor morphology. &
Gamma rays and neutrinos &
Galactic diffuse emission, large-scale instrumental acceptance,
cosmic-ray contamination, and halo normalization. &
Extragalactic diffuse emission, Galactic-center morphology, and
cross-correlation analyses. \\

Sun &
Precisely known moving source direction and very low directional
background for high-energy neutrinos; seven years of data were used in
a low-mass IceCube analysis. &
Neutrinos; gamma rays in secluded or long-lived mediator scenarios &
Capture rate, scattering cross section, capture-annihilation
equilibrium, solar composition, and solar-atmospheric backgrounds. &
Direct-detection constraints and Galactic-halo searches evaluated
within the same particle-physics model. \\

Extra-\newline galactic diffuse sky &
Very large cosmological volume and sensitivity to cumulative emission
from unresolved DM halos; Fermi-LAT measured the isotropic spectrum
over \(\sim100\) MeV--820 GeV. &
Gamma rays and high-energy neutrinos &
Unresolved astrophysical populations, cosmological foreground
separation, halo mass function, concentration relation, and
substructure. &
Angular cross-correlation with large-scale-structure tracers and
energy-spectrum decomposition. \\

M31 and nearby galaxies &
Independent halo systems with resolvable morphology and intermediate
angular extent between dwarfs and the Milky Way. &
Gamma rays; X-rays or line searches for suitable decay models &
Foreground subtraction, unresolved stellar sources, cosmic rays, and
host-halo modeling. &
Scaling tests against Milky Way, dwarf, and cluster targets. \\

Nearby galaxy groups &
Favorable compromise between halo mass, distance, and angular extent;
useful particularly for decay searches. &
Gamma rays and X-ray/line searches &
Group membership, diffuse astrophysical emission, and halo-mass
uncertainties. &
Comparison with individual galaxies and galaxy clusters. \\

Cosmo-\newline logical large-scale structure &
Spatial correlation of a DM signal with independently mapped matter
distributions provides information beyond the mean diffuse intensity. &
Gamma rays and neutrinos &
Bias modeling, tracer redshift distributions, unresolved source populations, and nonlinear structure formation. &
Cross-correlations using galaxy catalogs, weak lensing, or other large-scale-structure tracers. \\

CMB / early Universe &
Sensitive to energy injection over cosmological epochs rather than to an individual halo; particularly powerful for annihilation models with substantial energy deposition at recombination. &
CMB temperature and polarization anisotropies &
Energy-deposition efficiency, cosmological parameter degeneracies, and velocity dependence of the annihilation process. &
Late-time gamma-ray, neutrino, and dwarf-galaxy searches. \\

Extra-\newline galactic line searches &
Redshift-broadened spectral signatures integrate contributions from cosmological DM distributions and can probe very long lifetimes. &

X-rays and gamma rays, depending on DM mass &
Instrumental lines, astrophysical atomic features, continuum modeling, and cosmological redshift integration. &
Galactic center, Milky Way halo, clusters, and nearby galaxies. \\

\midrule

\multicolumn{5}{@{}p{\columnwidth}@{}}{%
\footnotesize
\textit{Note.} No target simultaneously maximizes signal intensity, background cleanliness, spatial information, and astrophysical robustness. The optimal target therefore depends on the annihilation or decay hypothesis, DM mass and final state, instrument response, region of interest, and how astrophysical nuisance parameters are treated. Cross-target consistency provides an important test because a genuine DM signal should scale with the corresponding \(J\)- or \(D\)-factor and reproduce the expected spectral and spatial behavior across independent environments.
} \\

\end{longtable}

\normalsize

\section{Gamma Rays: The Most Mature and Most Model-Entangled Channel}\label{gamma-rays-the-most-mature-and-most-model-entangled-channel}

Gamma rays are central to indirect DM searches because they
approximately preserve the source direction and can produce continuum, endpoint, or line-like spectral signatures. Quantitative comparisons must nevertheless retain the stated DM mass, annihilation channel, confidence level, halo profile, and observing strategy; visually adjacent exclusion curves do not necessarily constrain the same hypotheses.

Fermi-LAT dwarf-spheroidal analyses provide some of the most robust GeV-scale constraints. The six-year analysis of 15 dwarfs found no significant gamma-ray emission and excluded the canonical thermal-relic cross section for quark and \ensuremath{\tau}-lepton channels below approximately 100~GeV \cite{ref8}. A subsequent 14-year analysis likewise found no significant combined signal, although several individual dwarfs showed local excesses above 2\ensuremath{\sigma} \cite{ref9}. At TeV energies, the H.E.S.S. Inner Galaxy
Survey obtained limits of 3.7\ensuremath{\times}10\textsuperscript{-26} cm\textsuperscript{3} s\textsuperscript{-1} at \(m_{\chi}\) = 1.5~TeV for W\textsuperscript{+}W\textsuperscript{-} and
1.2 \ensuremath{\times} 10\textsuperscript{-26} cm\textsuperscript{3} s\textsuperscript{-1} at 0.7~TeV for \ensuremath{\tau}\textsuperscript{+}\ensuremath{\tau}\textsuperscript{-}, assuming an Einasto profile \cite{ref43}. These results are complementary rather than directly interchangeable: dwarf analyses propagate stellar-kinematic \(J\)-factor uncertainties, whereas Galactic-center limits depend strongly on the assumed inner halo profile.

Line searches offer greater spectral specificity but remain sensitive to instrumental systematics and trial factors. MAGIC obtained 95\% confidence limits on \ensuremath{\chi}\ensuremath{\chi} \ensuremath{\to} \ensuremath{\gamma}\ensuremath{\gamma} of approximately 5\ensuremath{\times}10\textsuperscript{-28}~cm\textsuperscript{3} s\textsuperscript{-1} at 1 TeV
and 1 \ensuremath{\times} 10\textsuperscript{-25} cm\textsuperscript{3} s\textsuperscript{-1} at 100 TeV, using 223 h of Galactic-center observations under a cuspy halo assumption \cite{ref44}. A 15-year Fermi-LAT analysis found no significant spectral line in the GeV regime \cite{ref47}.
Any candidate feature must also pass detector-frame,
global-significance, and spatial-morphology tests. A 2026 H.E.S.S. line search based on 546 h of Inner Galaxy Survey data likewise reported no significant signal and reached approximately 2.3 \ensuremath{\times} 10\textsuperscript{-28} cm\textsuperscript{3} s\textsuperscript{-1} at 1~TeV for an Einasto profile.

At higher masses, HAWC used 2,865 days of Galactic-center observations to probe WIMP masses from 1 TeV to 10 PeV, reaching limits of order 10\textsuperscript{-24} cm\textsuperscript{3} s\textsuperscript{-1} for the \ensuremath{\tau}\textsuperscript{+}\ensuremath{\tau}\textsuperscript{-} channel under an Einasto profile \cite{ref51}.
HAWC and LHAASO diffuse measurements extend sensitivity further into the TeV--PeV regime \cite{ref37}, where electroweak showering, secondary emission, gamma-ray absorption, and halo normalization increasingly
shape the inferred constraint.

Instrumental complementarity is equally important. Space-based
pair-conversion telescopes provide wide-field, long-duration surveys; X-ray, coded-mask, and Compton instruments extend coverage into the keV--MeV range; imaging atmospheric Cherenkov telescopes provide large effective areas at TeV energies, and wide-field particle or water-Cherenkov arrays offer high duty cycles and broad sky coverage. Their sensitivities depend not only on exposure but also on spectral and angular response, background environment, and source-model systematics
\cite{ref38,ref52,ref53}.

For example, Fermi-LAT enables long-baseline population studies but is limited at low energies by a broad point-spread function and by diffuse-emission modeling. INTEGRAL/SPI offers strong MeV line sensitivity but faces instrumental-background and coded-mask limitations
for extended emission. XMM-Newton provides a deep archival resource for decay searches, while NuSTAR benefits from focusing optics but must control stray-light and aperture backgrounds. Ground-based Cherenkov telescopes achieve strong event reconstruction over relatively narrow fields and often rely on nearby control regions that can partially
subtract extended DM emission. Wide-field arrays instead favor extended structures and survey searches but require precise control of hadronic backgrounds and large-scale acceptance.

The MeV regime remains both an instrumental and a theoretical frontier. Light DM annihilation may produce correlated in-flight annihilation, bremsstrahlung, and 511-keV emission, while primordial black-hole evaporation produces linked photon, lepton, and neutrino spectra. Model these components jointly rather than as independent signals.
Improved Compton imaging and spectral resolution may help distinguish halo-like emission from gas-correlated and instrumental backgrounds, provided analyses include realistic activation backgrounds and extended-source reconstruction.

Joint analyses should therefore convolve a common physical signal model with each instrument response and include facility-specific nuisance parameters. Compare published gamma-ray results using their stated channels, masses, halo assumptions, response characteristics, and dominant systematics, not exposure alone. The relevant comparison is the full exposure--response--background likelihood, not observing time or effective area in isolation.

\subsection{Dwarf Spheroidal Galaxies}\label{dwarf-spheroidal-galaxies}

Dwarfs remain the benchmark for robust GeV annihilation limits. Table~\ref{tab:dwarf_gamma_searches} traces the sequence of Fermi-LAT analyses. Pass 8, longer exposure, additional satellites, and improved interstellar modeling increased sensitivity, but the core logic remained a joint likelihood in which each dwarf's J factor is constrained by stellar
kinematics. The 14-year legacy analysis found no population-level excess and reinforced the importance of target selection and J-factor treatment
\cite{ref8,ref9,ref54,ref55}. Its scientific value exceeds that
of a single exclusion curve: it provides a clean reference against which
interpretations of the Galactic center, M31, clusters, and
charged-particle anomalies can be tested.

Table~\ref{tab:dwarf_gamma_searches} compares the published dwarf-galaxy searches, their target samples, representative results, and principal analysis limitations.

\setlength{\LTcapwidth}{\columnwidth}
\setlength{\tabcolsep}{2pt}
\renewcommand{\arraystretch}{1.15}
\setlength{\emergencystretch}{2em}

\footnotesize

\begin{longtable}[]{@{}
  >{\raggedright\arraybackslash}p{(\columnwidth - 8\tabcolsep) * \real{0.16}}
  >{\raggedright\arraybackslash}p{(\columnwidth - 8\tabcolsep) * \real{0.18}}
  >{\raggedright\arraybackslash}p{(\columnwidth - 8\tabcolsep) * \real{0.22}}
  >{\raggedright\arraybackslash}p{(\columnwidth - 8\tabcolsep) * \real{0.22}}
  >{\raggedright\arraybackslash}p{(\columnwidth - 8\tabcolsep) * \real{0.22}}
@{}}

\caption{Dwarf-galaxy gamma-ray searches with study-level exposure. Sources: \cite{ref8,ref9,ref54,ref55,ref60,ref61}.}
\label{tab:dwarf_gamma_searches} \\

\toprule
\textbf{Study} &
\textbf{Data / targets} &
\textbf{Mass / energy scope} &
\textbf{Reported result} &
\textbf{Confidence / systematic} \\
\midrule
\endfirsthead

\toprule
\textbf{Study} &
\textbf{Data / targets} &
\textbf{Mass / energy scope} &
\textbf{Reported result} &
\textbf{Confidence / systematic} \\
\midrule
\endhead

\midrule
\multicolumn{5}{r}{\footnotesize Continued on next page} \\
\endfoot

\bottomrule
\endlastfoot


Fermi-LAT 2015 &
6 years Pass 8; 15 dwarfs &
Low-GeV to TeV WIMPs &
No significant excess; below the thermal scale for selected channels
below about 100 GeV &
95\% limits; \(J\)-factor nuisance \\

Geringer-Sameth et al. 2015 &
Independent dwarf event analysis &
Channel-dependent WIMP scan &
No discovery; independent likelihood constraints &
Event weighting and \(J\)-factors \\

Fermi new satellites 2017 &
DES-era satellite candidates &
GeV WIMP scan &
No robust population signal &
Target confirmation and spectroscopy \\

VERITAS 2017 &
Joint northern dwarf observations &
About 100 GeV--100 TeV &
No significant signal &
95\% limits; exposure/\(J\) hierarchy \\

Multi-observatory 2022 &
Fermi, HAWC, H.E.S.S., MAGIC, VERITAS; MAGIC total 354.3 h &
5 GeV--100 TeV class &
No significant combined signal &
95\% limits; harmonized spectra/\(J\)-factors \\

Fermi-LAT 2024 &
\(>14\) years; updated census &
GeV--TeV channels &
Seven local excesses; no significant combined DM signal &
Local \(\sim2\)--\(3\sigma\) structure; \(J\)-selection \\

\midrule

\multicolumn{5}{@{}p{\columnwidth}@{}}{%
\footnotesize
\textit{Note.} The thermal-scale statement is channel- and
mass-dependent. Exposure values are not converted to a common
effective exposure.
} \\

\end{longtable}

\normalsize

The dwarf record shows how exposure gains can be limited by target knowledge. The evolution of dwarf analyses reveals that more targets do not guarantee stronger or more reliable limits. Newly discovered ultrafaints may have large, estimated J factors but few member stars, possible tidal disturbance, and uncertain dynamical equilibrium.
Including them can improve a median limit while enlarging vulnerability to a single prior. Classical dwarfs often have lower nominal J factors but better constrained kinematics. The correct weighting is supplied by the full likelihood, not a ranked list. Long Fermi-LAT exposure also enters a regime in which diffuse-model and source-catalog choices can matter as much as counting statistics. Future progress depends on optical spectroscopy, membership modeling, binary corrections, and transparent treatment of non-detections in the stellar sample. A valuable legacy product would connect the gamma-ray likelihood directly to versioned stellar-kinematic posteriors, allowing the particle constraint to update when a dwarf's dynamical model changes without repeating the full photon analysis. For IACT dwarfs, observing conditions couple to target selection. A high-J northern dwarf observed at favorable zenith angles may outperform a nominally brighter target with a higher threshold. Energy threshold matters because continuum spectra fall rapidly near the particle mass and because low-energy overlap with Fermi-LAT anchors the combination. The small field of view also makes extension and background estimation relevant for nearby dwarfs. Cross-observatory combinations are strongest when each experiment contributes a likelihood over reconstructed energy and direction. If only upper limits are combined, differences in nuisance profiling, event selection, and confidence construction cannot be recovered. CTAO can improve this landscape through standardized analysis products and a portfolio that includes both deep established dwarfs and flexible follow-up of newly discovered satellites. The most valuable new dwarf may be the one with moderate J and excellent spectroscopic certainty, not the one with the largest preliminary central value.

IACT dwarf programs are assessed in Table~\ref{tab:dwarf_gamma_searches}. Their greatest
leverage lies at masses beyond the Fermi-LAT optimum, especially for hard channels. Deep observations of a single favorable dwarf can be competitive, but a target's nominal J factor is not sufficient: uncertainty, angular extent, zenith-angle distribution, and exposure determine its actual weight. Cross-observatory combinations \cite{ref60} are a methodological advance because they cover a broad mass range, yet they must harmonize channel spectra, halo priors, and the meaning of a 95\% limit. Combining plotted curves is not equivalent to combining likelihoods.

\subsection{Galactic Center Continuum}\label{galactic-center-continuum}

The Galactic center offers a strong expected WIMP signal but also presents one of the most complex gamma-ray foreground environments. The continuum analysis choices
that move a result are arrayed in Table~\ref{tab:gc_continuum_evidence}. The inner profile sets
the signal amplitude; the ROI changes both signal concentration and
foreground covariance; diffuse templates encode gas, cosmic-ray sources,
and inverse-Compton emission; point-source masks redistribute photons
between resolved and diffuse components; and channel assumptions connect
an observed spectrum to mass. Treat these choices as a model ensemble or
a hierarchical nuisance problem rather than a small set of decorative
alternative curves.

\setlength{\LTcapwidth}{\columnwidth}
\setlength{\tabcolsep}{2pt}
\renewcommand{\arraystretch}{1.15}
\setlength{\emergencystretch}{2em}

\footnotesize

\begin{longtable}[]{@{}
  >{\raggedright\arraybackslash}p{(\columnwidth - 8\tabcolsep) * \real{0.16}}
  >{\raggedright\arraybackslash}p{(\columnwidth - 8\tabcolsep) * \real{0.18}}
  >{\raggedright\arraybackslash}p{(\columnwidth - 8\tabcolsep) * \real{0.22}}
  >{\raggedright\arraybackslash}p{(\columnwidth - 8\tabcolsep) * \real{0.23}}
  >{\raggedright\arraybackslash}p{(\columnwidth - 8\tabcolsep) * \real{0.21}}
@{}}

\caption{Galactic-center continuum evidence and sensitivity-moving choices. Sources: \cite{ref43,ref62,ref63,ref64,ref65,ref66,ref67}.}
\label{tab:gc_continuum_evidence} \\

\toprule
\textbf{Study / choice} &
\textbf{Data scale} &
\textbf{Quantitative feature} &
\textbf{Alternative / control} &
\textbf{Implication} \\
\midrule
\endfirsthead

\toprule
\textbf{Study / choice} &
\textbf{Data scale} &
\textbf{Quantitative feature} &
\textbf{Alternative / control} &
\textbf{Implication} \\
\midrule
\endhead

\midrule
\multicolumn{5}{r}{\footnotesize Continued on next page} \\
\endfoot

\bottomrule
\endlastfoot


Daylan et al. &
Inner Galaxy, GeV photons &
Spectrum peaks at a few GeV; roughly 30--50~GeV \(b\)-quark scale &
Diffuse templates and point sources &
DM-compatible, but not source-unique \\

Calore et al. &
Diffuse-model ensemble &
Foreground covariance broadens the allowed spectrum &
Many interstellar-emission realizations &
Systematic covariance must enter fits \\

Fermi-LAT 2017 &
Multiple ROIs and diffuse models &
Extended excess retained under model variants &
Bulge/source templates compete &
Residual persistence exceeds origin specificity \\

Di Mauro 2021 &
11 years of Fermi-LAT data &
GCE characterized over a broad inner-Galaxy region &
Diffuse and source-model variants &
Long exposure does not eliminate model dependence \\

NPTF injection tests &
\(10^{\circ}\) ROI in related analysis &
Injected smooth signals can be misassigned &
Signal-model mismatch &
Point-source preference is not automatically decisive \\

Foreground marginalization &
Data-driven harmonic modes &
Point-source preference survives tested alternatives &
Improved diffuse description &
Published methods reach opposing conclusions \\

H.E.S.S. 2022 &
2014--2020 TeV Galactic-center survey &
Best Einasto anchors:
\(3.7\times10^{-26}\) at 1.5~TeV and
\(1.2\times10^{-26}\) at 0.7~TeV &
NFW/Einasto profile tests &
Strong TeV bound conditional on a cuspy halo profile \\

\midrule

\multicolumn{5}{@{}p{\columnwidth}@{}}{%
\footnotesize
\textit{Note.} Cross sections are in
\(\mathrm{cm^{3}\,s^{-1}}\).
The H.E.S.S. values preserve the benchmark masses reported in the
abstract; channel labels remain in the source record.
} \\

\end{longtable}

\normalsize

The Galactic center concentrates both signal and model risk. The
Galactic-center continuum problem reflects model uncertainty that a
single nuisance amplitude cannot capture. Gas-correlated pion emission,
bremsstrahlung, inverse-Compton light, unresolved sources, and
large-scale structures have distinct spectra and morphologies, and
errors in cosmic-ray source distributions propagate across them.
Changing the ROI alters which components are constrained internally.
Masking the plane reduces foreground intensity but also discards the
region with the greatest signal leverage. A flexible diffuse model can
absorb a true halo component; a rigid one can create a false excess. The
appropriate analysis compares families with controlled complexity and
validates signal recovery under mis-specified foregrounds. Spatial
regions not used to fit the dark component should serve as predictive
checks. Reporting the stability of mass and cross section, rather than
only the test statistic, is especially informative: a genuine particle
spectrum should not migrate dramatically when a plausible gas map or
source mask is changed.

\begin{figure}[htbp]
\centering
\includegraphics[width=0.95\linewidth]{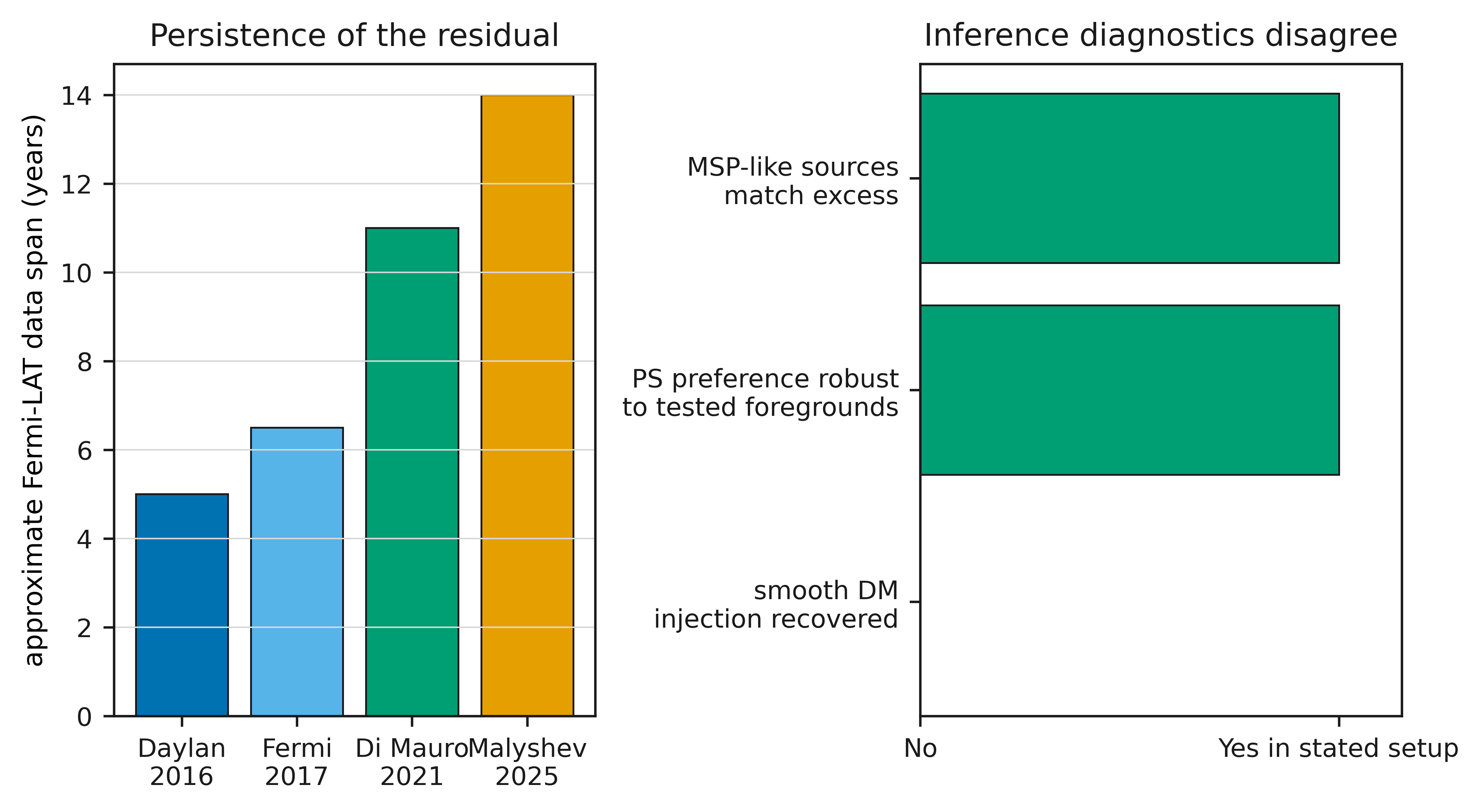}
\caption{Quantitative audit of the Galactic-center excess. The panels separate measured or reported landmarks from inference diagnostics affected by diffuse and source-model choices. Sources: \cite{ref62,ref63,ref64,ref65,ref66,ref67}.}
\label{fig:gc-excess}
\end{figure}

Fig.~\ref{fig:gc-excess} separates the persistence of the Galactic-center residual
from the analysis diagnostics used to infer its origin. The left panel
records approximate Fermi-LAT data spans for representative studies,
while the right panel indicates whether specific injection and source
tests succeeded under their stated setups. This format makes clear that
a long-lived residual and a unique DM explanation are distinct
evidential claims.

The GeV excess has persisted across many event selections and diffuse
models \cite{ref62,ref63,ref64,ref65,ref68,ref69,ref70}. Its status as a residual is more secure than its origin. DM
can reproduce a broad, approximately spherical component with a cross
section near the thermal scale in some channels. Millisecond pulsars can
reproduce the spectrum and naturally trace an old stellar population;
cosmic-ray outbursts and mismodeled inverse-Compton emission add
flexibility. The decisive question is not whether a dark template
improves a single baseline likelihood, but whether its inferred
morphology, spectrum, and normalization remain stable when source
populations and diffuse uncertainties are modeled with comparable
flexibility. The contrasting observational landmarks and modeling tests relevant to the Galactic-center excess are illustrated in Fig.~\ref{fig:gc-excess}.

\subsection{Lines, Endpoints, and Trials}\label{lines-endpoints-and-trials}

Gamma-ray lines are often called smoking guns because conventional
sources rarely generate narrow features at DM energies. That slogan
should not obscure the analysis problem. Line-search analyses require particular attention to control samples, calibration artifacts, and the global significance of scanned spectral features. Fermi-LAT, H.E.S.S., and MAGIC have
conducted broad scans with no robust Galactic line \cite{ref45,ref46,ref47,ref56,ref71,ref72}. Energy dispersion converts a
monochromatic injection into an instrument-dependent feature, and
control regions and event-quality partitions can reveal line-like
calibration artifacts. The mass scan, multiple ROIs, multiple line
morphologies, and analysis variants contribute a look-elsewhere penalty. Table~\ref{tab:gamma_line_searches} compares representative gamma-ray line searches while retaining their exposure, target, and statistical qualifications.

\setlength{\LTcapwidth}{\columnwidth}
\setlength{\tabcolsep}{2pt}
\renewcommand{\arraystretch}{1.15}
\setlength{\emergencystretch}{2em}

\footnotesize

\begin{longtable}[]{@{}
  >{\raggedright\arraybackslash}p{(\columnwidth - 8\tabcolsep) * \real{0.16}}
  >{\raggedright\arraybackslash}p{(\columnwidth - 8\tabcolsep) * \real{0.19}}
  >{\raggedright\arraybackslash}p{(\columnwidth - 8\tabcolsep) * \real{0.21}}
  >{\raggedright\arraybackslash}p{(\columnwidth - 8\tabcolsep) * \real{0.22}}
  >{\raggedright\arraybackslash}p{(\columnwidth - 8\tabcolsep) * \real{0.22}}
@{}}

\caption{Gamma-ray line searches and trials-aware numerical results. Sources: \cite{ref11,ref44,ref46,ref47,ref71,ref72}.}
\label{tab:gamma_line_searches} \\

\toprule
\textbf{Search} &
\textbf{Exposure / range} &
\textbf{Target} &
\textbf{Published result} &
\textbf{Qualifier} \\
\midrule
\endfirsthead

\toprule
\textbf{Search} &
\textbf{Exposure / range} &
\textbf{Target} &
\textbf{Published result} &
\textbf{Qualifier} \\
\midrule
\endhead

\midrule
\multicolumn{5}{r}{\footnotesize Continued on next page} \\
\endfoot

\bottomrule
\endlastfoot


Fermi-LAT 2015 &
5.8 years; 200 MeV--500~GeV class &
Multiple Galactic ROIs &
No globally significant line &
Energy dispersion and mass/ROI trials \\

H.E.S.S. 2018 &
10 years; about 254 h &
Inner Galactic halo &
No line; strongest TeV-era limits &
95\% CL; Einasto/cusp conditional \\

MAGIC 2023 &
223 h; up to 100~TeV &
Galactic center &
\(\langle\sigma v\rangle < 5\times10^{-28}\,
\mathrm{cm^{3}\,s^{-1}}\) at 1~TeV;
\(\langle\sigma v\rangle < 1\times10^{-25}\,
\mathrm{cm^{3}\,s^{-1}}\) at 100~TeV &
95\% CL; two-photon line, cuspy profile \\

Fermi 14-year GC &
14 years; broad GeV scan &
Galactic center &
No robust line &
Local/global scan and diffuse control \\

Fermi 15-year &
15 years &
Galactic ROIs &
No line; single- and double-line constraints &
Higgs-portal interpretation is model dependent \\

Cluster stack 2026 &
Virgo, Fornax, Ophiuchus plus cluster sample &
About 43 GeV &
Published line-like excess &
No corresponding inner-Galaxy line; replication pending \\

\midrule

\multicolumn{5}{@{}p{\columnwidth}@{}}{%
\footnotesize
\textit{Note.} Cross sections are in
\(\mathrm{cm^{3}\,s^{-1}}\).
Values use each study's halo convention and must not be joined as one
recast curve.
} \\

\end{longtable}

\normalsize

A line is specific in energy but not necessarily in origin. Line
searches benefit from a localized spectral hypothesis, but this
localization concentrates instrumental risk. Energy reconstruction
tails, gain drift, event-quality dependence, and imperfectly modeled
background lines can all produce narrow residuals. A credible search
should show the feature under independent calibrations and event
partitions, quantify its width relative to the response, and examine
control regions with similar incidence angles or detector coordinates.
Target stacking adds a second layer: weights and the target list should
be declared before the line energy is selected, or the selection becomes
part of the trials factor. The 43-GeV cluster result is particularly
instructive because the missing inner-Galaxy counterpart is not a minor
tension; under standard annihilation, it challenges the target scaling
that made the feature interesting. A noncanonical model could evade that
expectation, but its added freedom must generate a new prediction rather
than merely excuse the absent signal.

The published 2026 report of an approximately 43-GeV line in stacked
Virgo, Fornax, and Ophiuchus data is an important stress test \cite{ref11}.
A line-like feature across several clusters is intriguing, but its
absence from the inner Galaxy disfavors canonical annihilation under
standard halo scaling. This is scientifically productive
counterevidence: it narrows the space of viable interpretations toward
an instrumental, astrophysical, target-selection, or noncanonical
dark-sector explanation. Independent analyses, predefined target sets,
control-energy scans, and consistency with continuum limits are required
before the feature can move up the evidence ladder.

\subsection{Clusters, the Isotropic Sky, and Nearby Galaxies}\label{clusters-the-isotropic-sky-and-nearby-galaxies}

Cluster, extragalactic, and unassociated-source searches are summarized
in Table~\ref{tab:clusters_diffuse_subhalos}. Sixteen years of Fermi-LAT observations of nearby
clusters show no \textgreater5\ensuremath{\sigma} continuum population signal, though some
regions exhibit elevated test statistics \cite{ref45}. Cluster searches are
limited by uncertain subhalo boosts and astrophysical backgrounds from
AGN and cosmic-ray interactions; stacking reduces statistical noise but
not shared modeling or selection biases.

The isotropic gamma-ray background constrains annihilation and decay
over cosmic history, but its interpretation depends strongly on
unresolved blazars, star-forming galaxies, and misaligned active nuclei
\cite{ref13,ref48,ref73,ref74}. Anisotropy and
cross-correlations with galaxy catalogs can help separate these
components by their distinct redshift and bias distributions, provided
that catalog selection and redshift kernels are modeled consistently.

Unassociated-source searches test the prediction that Galactic subhalos
may appear as steady gamma-ray sources without conventional
counterparts. Candidate selection typically relies on steadiness,
spectral shape, latitude, extension, and the absence of counterparts,
but each criterion carries astrophysical false positives and depends on
survey efficiency and diffuse-background modeling. Therefore,
population-level agreement in source counts, flux, latitude, and
extension is more informative than a single high-scoring candidate.
Machine-learning methods can prioritize follow-up, but their
classifications require representative training samples, calibrated
priors, and treatment of unknown source classes.

M31 and other nearby galaxies provide an intermediate scale between
Galactic and cosmological searches, offering independent halo geometries
and resolved morphologies. Their interpretation remains limited by
cosmic-ray emission and foreground subtraction. A consistent DM model
should describe dwarfs, clusters, nearby galaxies, and diffuse
backgrounds without requiring target-specific cross sections or
arbitrary boost factors.

The MeV band provides complementary sensitivity to light DM, positron
injection, nuclear backgrounds, and evaporating primordial black holes.
INTEGRAL constrains sub-GeV annihilation via diffuse emission and the
511~keV environment \cite{ref38}, while COSI- and e-ASTROGAM-class
instruments promise improved imaging and continuum sensitivity \cite{ref56,ref57}. Their performance will depend on activation-line calibration,
response tails, and separation of gas-correlated emission from halo-like
morphology.

Representative gamma-ray constraints should be compared by benchmark sensitivity and observing scale rather than by an interpolated sensitivity curve. Dwarf galaxies provide robust GeV benchmarks; imaging
atmospheric Cherenkov telescopes in the inner Galaxy offer strong TeV
sensitivity under cuspy halo assumptions; and wide-field arrays extend
searches to higher masses. Comparisons must retain each
analysis's published spectrum, halo model, instrument
response, and nuisance assumptions. Table~\ref{tab:clusters_diffuse_subhalos} brings together the complementary cluster, diffuse-sky, and subhalo results and their leading astrophysical limitations.

\setlength{\LTcapwidth}{\columnwidth}
\setlength{\tabcolsep}{2pt}
\renewcommand{\arraystretch}{1.15}
\setlength{\emergencystretch}{2em}

\footnotesize

\begin{longtable}[]{@{}
  >{\raggedright\arraybackslash}p{(\columnwidth - 8\tabcolsep) * \real{0.16}}
  >{\raggedright\arraybackslash}p{(\columnwidth - 8\tabcolsep) * \real{0.18}}
  >{\raggedright\arraybackslash}p{(\columnwidth - 8\tabcolsep) * \real{0.22}}
  >{\raggedright\arraybackslash}p{(\columnwidth - 8\tabcolsep) * \real{0.22}}
  >{\raggedright\arraybackslash}p{(\columnwidth - 8\tabcolsep) * \real{0.22}}
@{}}

\caption{Clusters, diffuse gamma rays, and catalog-selected subhalo searches. Sources: \cite{ref11,ref45,ref48,ref49,ref51,ref73,ref75}.}
\label{tab:clusters_diffuse_subhalos} \\

\toprule
\textbf{Study / observable} &
\textbf{Sample / range} &
\textbf{Numerical outcome} &
\textbf{Background / control} &
\textbf{Evidence status} \\
\midrule
\endfirsthead

\toprule
\textbf{Study / observable} &
\textbf{Sample / range} &
\textbf{Numerical outcome} &
\textbf{Background / control} &
\textbf{Evidence status} \\
\midrule
\endhead

\midrule
\multicolumn{5}{r}{\footnotesize Continued on next page} \\
\endfoot

\bottomrule
\endlastfoot


Fermi clusters 2025 &
16 clusters; 16 years &
No \(>5\sigma\) DM signal; Coma and Perseus show peak
\(\mathrm{TS}>16\); Perseus reaches
\(\mathrm{TS}\approx20\) at \(m_{\chi}\approx38\) GeV in the
\(\tau\) channel. &
Time evolution and dwarf conflict &
Target-specific residuals \\

43-GeV line &
Cluster stacking &
Feature near 43 GeV &
Inner-Galaxy non-detection and target selection &
Published anomaly, unreplicated \\

Isotropic spectrum &
100 MeV--820 GeV &
Measured diffuse intensity spectrum &
Blazars, star-forming galaxies, misaligned AGN &
Input to decay/annihilation limits \\

Resolved EGB &
\(>50\) GeV &
86\% \((+16/-14\%)\) attributed to point sources &
Population extrapolation &
Little residual room, but uncertainty is material \\

M31 outer halo &
Extended outer-halo ROI &
Residual compatible with several source models &
Milky Way foreground and cosmic rays &
Not a unique DM detection \\

2HWC catalog &
507 days; 39 sources &
High-duty-cycle TeV catalog &
Selection function and source association &
Subhalo candidate input, not DM evidence \\

HAWC GC 2026 &
2,865 days; \(\pm9^{\circ}\); 1 TeV--10 PeV &
No significant excess; strongest limits of order
\(10^{-24}\) &
Three profiles and three channels &
95\% high-mass limits \\

\midrule

\multicolumn{5}{@{}p{\columnwidth}@{}}{%
\footnotesize
\textit{Note.} TS = test statistic. The relation between TS and
\(\sigma\) depends on the number of degrees of freedom, parameter
boundaries, and trials.
} \\

\end{longtable}

\normalsize

\section{Charged Cosmic Rays: Precision Spectra, Nonlocal Inference}\label{charged-cosmic-rays-precision-spectra-nonlocal-inference}

The measured positron excess is well established, although its origin remains uncertain. PAMELA established a rising positron fraction from 1.5--100~GeV \cite{ref76}. AMS-02 subsequently based its high-energy study on 1.9
million positrons extending to 1 TeV, measured a sharp flux drop above
\(284_{-64}^{+91}\)~GeV, and inferred a source cutoff of
\(810_{-180}^{+310}\)~GeV with significance above \(4\sigma\)
\cite{ref77}. DAMPE measured the combined electron and positron spectrum
from 25 GeV to 4.6 TeV and reported a direct break near 0.9~TeV
\cite{ref78}. Although spatially distributed DM sources can reproduce the
observed smooth high-energy spectra, conventional astrophysical sources,
particularly pulsars combined with radiative energy losses during
propagation, can produce similar spectral features and cutoffs.
Consequently, spectral agreement alone does not uniquely imply a DM
interpretation. Greater discriminating power would come from correlated
observables, such as a characteristic anisotropy or an associated
gamma-ray counterpart, whose absence currently limits the ability to
distinguish DM scenarios from astrophysical explanations.

Antiprotons illustrate why significance should be reported as a property
of an analysis specification. AMS-02 measured approximately 349,000
antiprotons over 0.5--450~GV, with the antiproton-to-proton ratio
approximately flat above 60 GV \cite{ref79}. Published DM fits have assigned
local significances ranging from negligible to several
\(\sigma\) depending on production cross sections, transport
flexibility, solar modulation, and covariance \cite{ref80,ref81,ref82,ref83,ref84}. The comparison in Table~\ref{tab:antiproton_evidence} keeps
those assumptions visible rather than averaging incompatible
significances.

Charged cosmic rays provide high-statistics spectra and particle
identification but surrender source direction to magnetic diffusion. The principal measurements span complementary observables. PAMELA
established the positron anomaly, AMS-02 transformed it and the
antiproton spectrum into precision measurements, Fermi-LAT and DAMPE
extended calorimetric (e\textsuperscript{-}+e\textsuperscript{+})
spectra, Voyager sampled the local interstellar medium beyond the
heliopause, and GAPS targets low-energy antinuclei. These instruments do
not observe the same quantity: rigidity, kinetic energy, charge-sign
separation, acceptance, and time coverage must be matched before
combining spectra \cite{ref76,ref77,ref78,ref79,ref85,ref86,ref87}.

For charged particles, propagation is part of the signal model.
Instrument complementarity is useful only if native observables and
covariance are preserved: AMS-02 measures charge-separated rigidity
spectra, whereas calorimetric experiments generally measure energy
spectra, often without charge separation. Conversions among rigidity,
kinetic energy, and total energy therefore introduce species-dependent
effects, so joint analyses should convolve a common physical model with
each detector response rather than combine central-value spectra.

Recent AMS-02 measurements underscore the importance of heliospheric
transport. Full-solar-cycle data show distinct time dependence for
antiprotons, protons, electrons, positrons, and nuclei, demonstrating
that solar modulation depends on charge, spectral shape, and rigidity
\cite{ref90,ref91}. Below tens of GV, time averaging or simple
force-field treatments can therefore bias propagation and DM inference.
Recent analyses of the time-dependent AMS-02 data confirm that the
assumed solar-modulation prescription can materially affect inferred
Galactic transport parameters \cite{ref92}.

Voyager 1 provides complementary measurements of the low-energy,
very-local interstellar spectrum beyond the heliopause, but does not
resolve uncertainties in Galactic transport. Recent isotope measurements
and propagation studies show that nuclear-production cross sections and
the local gas and source distributions can significantly affect inferred
transport parameters \cite{ref93,ref94}. B/C remains a primary grammage
constraint, but degeneracies among diffusion normalization, halo size,
reacceleration, and convection require additional information from
radioactive isotopes, synchrotron emission, and gamma-ray emissivity.
These observables introduce their own uncertainties in nuclear cross
sections, gas, and magnetic fields.

The relevant uncertainties are also species-dependent. High-energy
electrons and positrons undergo rapid radiative losses and are sensitive
to the discrete local source population, whereas antiproton predictions
depend strongly on production and tertiary-interaction cross sections.
Recent precision studies emphasize that nuclear and antimatter
production cross sections are now a limiting systematic when extracting
propagation parameters from cosmic-ray data \cite{ref95}. GAPS adds an
orthogonal low-energy antinucleus measurement via exotic-atom
identification; its first 25-day Antarctic science flight was completed
in January 2026, providing a complementary approach to magnetic
spectrometers.

Propagation, solar modulation, production cross sections, and detector
systematics should therefore be fitted or marginalized jointly wherever
possible. Selecting a transport model to obtain an appealing residual
and then evaluating that residual with propagation parameters
artificially fixed inflates its apparent significance. Robust DM
searches instead require evaluating residual significance across the
allowed transport and cross-section posterior.

\subsection{Transport as Part of the Signal Model}\label{transport-as-part-of-the-signal-model}

Galactic transport is part of the predicted charged-particle signal, not
a correction applied after source modeling. In steady state, the
transport equation can be written as

\begin{equation}
\resizebox{0.98\linewidth}{!}{$\displaystyle
0 = q(x,p)
+ \nabla \cdot \left( D_{xx}\nabla\psi - V_{c}\psi \right)
+ \frac{\partial}{\partial p}
\left[
p^{2}D_{pp}
\frac{\partial}{\partial p}
\left( \frac{\psi}{p^{2}} \right)
\right]
- \frac{\partial}{\partial p}
\left[
\left(
\dot{p} - \frac{p}{3}\nabla \cdot V_{c}
\right)\psi
\right]
- \frac{\psi}{\tau_{\mathrm{frag}}}
- \frac{\psi}{\tau_{\mathrm{decay}}}
$}
\label{eq:transport}
\end{equation}

\noindent where \ensuremath{\psi} is the cosmic-ray density per unit momentum, q includes
conventional and DM sources, \(D_{xx}\) represents spatial diffusion,
\(V_{c}\) represents convection, \(D_{pp}\) represents diffusive
reacceleration, and the remaining terms account for continuous energy
losses, fragmentation, and radioactive decay. 

Secondary-to-primary ratios, such as B/C, constrain combinations of
propagation parameters rather than uniquely determining diffusion, halo
size, reacceleration, or convection \cite{ref82,ref84,ref88,ref89}. Radioactive isotopes provide additional leverage on the
residence time, but recent Voyager 1 isotope measurements show that
fragmentation and production cross sections remain significant
uncertainties \cite{ref94}. Moreover, detailed modeling of Voyager 1 data
shows that the local gas and source distributions can alter the inferred
low-energy propagation parameters even beyond the heliopause. Spatially
uniform diffusion and a fixed propagation halo should therefore be
regarded as useful baseline assumptions rather than uniquely established
descriptions \cite{ref93}.

Solar modulation must likewise be incorporated into the transport
likelihood. AMS-02 measurements over a full solar cycle show rigidity-,
spectral-, and charge-dependent modulation \cite{ref90,ref91}. A 2026
analysis using time-dependent AMS-02 data found systematic shifts of
approximately 10--15\% in reconstructed local interstellar spectra and
propagation parameters, depending on the modulation treatment,
substantially exceeding the percent-level precision of modern cosmic-ray
measurements \cite{ref92}. A single force-field potential may therefore be
adequate for some applications but is insufficient for precision
searches involving low-rigidity residuals.

Production cross sections are especially important for secondary
antiprotons and antinuclei because their astrophysical backgrounds are predicted from production and transport models rather than measured in a source-free control region. A
recent comprehensive review identifies nuclear and antimatter production
cross sections as a limiting systematic in precision cosmic-ray
propagation and indirect DM searches \cite{ref95}. Their uncertainties
should therefore enter the likelihood as nuisance parameters rather than
be fixed to a single production model.

Interpreting these measurements requires a joint transport likelihood. A DM source can be partially absorbed by
changes in diffusion, reacceleration, losses, or secondary production.
Fitting propagation once and freezing it understates uncertainty;
allowing arbitrary flexibility can erase genuine structure. A defensible
analysis jointly fits nuclei and the target species, incorporating
physically motivated spatial freedom, covariance, and out-of-sample
validation.

\subsection{Positrons and Electrons}\label{positrons-and-electrons}

The positron fraction exceeds the expectations of simple secondary
models and then softens at high energy \cite{ref76,ref77,ref86}.
Table~\ref{tab:positron_electron_anomalies} separates observations from interpretations. Leptophilic
DM can produce a smooth, hard excess, but s-wave models are constrained
by CMB energy deposition and gamma rays. Nearby pulsars naturally
provide primary positrons, and the HAWC detection of extended gamma-ray
halos around Geminga and Monogem revealed inhibited diffusion near
pulsars \cite{ref87}. That result did not settle the question: a
slow-diffusion bubble does not necessarily describe transport all the
way to Earth, and a population of sources can differ from two exemplars.

\setlength{\LTcapwidth}{\columnwidth}
\setlength{\tabcolsep}{2pt}
\renewcommand{\arraystretch}{1.15}
\setlength{\emergencystretch}{2em}

\footnotesize

\begin{longtable}[]{@{}
  >{\raggedright\arraybackslash}p{(\columnwidth - 8\tabcolsep) * \real{0.16}}
  >{\raggedright\arraybackslash}p{(\columnwidth - 8\tabcolsep) * \real{0.18}}
  >{\raggedright\arraybackslash}p{(\columnwidth - 8\tabcolsep) * \real{0.22}}
  >{\raggedright\arraybackslash}p{(\columnwidth - 8\tabcolsep) * \real{0.22}}
  >{\raggedright\arraybackslash}p{(\columnwidth - 8\tabcolsep) * \real{0.22}}
@{}}

\caption{Positron and electron anomalies with quantitative discriminants. Sources: \cite{ref6,ref76,ref77,ref78,ref83,ref87}.}
\label{tab:positron_electron_anomalies} \\

\toprule
\textbf{Observation} &
\textbf{Published scale} &
\textbf{DM interpretation} &
\textbf{Astrophysical interpretation} &
\textbf{Decisive test} \\
\midrule
\endfirsthead

\toprule
\textbf{Observation} &
\textbf{Published scale} &
\textbf{DM interpretation} &
\textbf{Astrophysical interpretation} &
\textbf{Decisive test} \\
\midrule
\endhead

\midrule
\multicolumn{5}{r}{\footnotesize Continued on next page} \\
\endfoot

\bottomrule
\endlastfoot


PAMELA rise &
1.5--100 GeV &
Leptophilic annihilation/decay &
Pulsars or secondary acceleration &
Anisotropy and gamma-ray counterpart \\

AMS source cutoff &
\(810^{+310}_{-180}\) GeV; \(>4\sigma\) finite cutoff &
Mass/cascade endpoint &
Finite pulsar age and energy losses &
Joint time-independent source model \\

AMS flux drop &
Above \(284^{+91}_{-64}\) GeV &
High-energy suppression &
Cooling and nearby-source stochasticity &
Charge-resolved spectrum and covariance \\

DAMPE break &
About 0.9 TeV &
Endpoint or cascade &
Propagation/source break &
Independent calorimetric confirmation \\

HAWC pulsar halos &
Extended TeV emission around Geminga and Monogem &
Constrains simple pulsar-delivery scenarios &
Implies inhibited local diffusion &
Two-zone transport and population synthesis \\

CMB counterpart &
\(p_{\rm ann}<3.2\times10^{-28}\) at 95\% confidence &
Restricts efficient \(s\)-wave leptonic models &
No effect on ordinary pulsar emission &
Velocity dependence or late production \\

\midrule

\multicolumn{5}{@{}p{\columnwidth}@{}}{%
\footnotesize
\textit{Note.} The cutoff significance is a statement about the AMS
source-term fit, not evidence for a DM origin.
} \\

\end{longtable}

\normalsize

The positron data require a high-energy source component, but they do not identify its origin. The positron evidence supports a new primary component more
strongly than any particular origin. DM can produce a smooth halo
contribution, but the required annihilation strength and leptonic
branching often generate inverse-Compton gamma rays and CMB deposition.
Pulsars provide a known acceleration mechanism and naturally introduce
cutoffs through age and cooling. The HAWC halos demonstrate unusual
transport around specific pulsars but do not uniquely determine the
diffusion coefficient along the full path to Earth. Source anisotropy is
potentially discriminating, though multiple sources and magnetic fields
can suppress it. The strongest analysis would fit the positron fraction,
the combined e\ensuremath{\pm} spectrum, resolved halo emission, the diffuse
inverse-Compton sky, and anisotropy within a single time-dependent
local-source model, then compare that predictive model with a halo
source. A better fit to the one-dimensional fraction alone is
insufficient because both hypotheses have enough spectral freedom to
follow broad curvature.

Neither the rising positron fraction nor its high-energy softening uniquely determines a particle mass.
Cooling, source age, distance, injection spectrum, and a pulsar
distribution can produce similar curvature. Anisotropy-resolved
gamma-ray halos and a consistent multiwavelength energy budget are more
discriminating than a better one-dimensional spectral fit.

\subsection{Antiprotons and Antinuclei}\label{antiprotons-and-antinuclei}

AMS-02 antiproton measurements prompted influential WIMP interpretations
near the tens-of-GeV scale \cite{ref80,ref81}. Table~\ref{tab:antiproton_evidence} explains
why the reported significance changed as analyses adopted new production
cross sections, propagation fits, solar modulation, and correlated
systematics. A residual that persists for one background family but not
under a plausible covariance model is a clue about the null hypothesis,
not a discovery. The most informative approach is a joint fit of
protons, helium, secondary nuclei, and antiprotons, with public
covariance and accelerator-informed cross-section nuisance parameters
\cite{ref82,ref84}.

\setlength{\LTcapwidth}{\columnwidth}
\setlength{\tabcolsep}{2pt}
\renewcommand{\arraystretch}{1.15}
\setlength{\emergencystretch}{2em}

\footnotesize

\begin{longtable}[]{@{}
  >{\raggedright\arraybackslash}p{(\columnwidth - 8\tabcolsep) * \real{0.16}}
  >{\raggedright\arraybackslash}p{(\columnwidth - 8\tabcolsep) * \real{0.18}}
  >{\raggedright\arraybackslash}p{(\columnwidth - 8\tabcolsep) * \real{0.22}}
  >{\raggedright\arraybackslash}p{(\columnwidth - 8\tabcolsep) * \real{0.22}}
  >{\raggedright\arraybackslash}p{(\columnwidth - 8\tabcolsep) * \real{0.22}}
@{}}

\caption{Antiproton evidence and analysis-dependent significance. Sources: \cite{ref79,ref80,ref81,ref82,ref84}.}
\label{tab:antiproton_evidence} \\

\toprule
\textbf{Study / component} &
\textbf{Data scale} &
\textbf{Quantitative finding} &
\textbf{Null-model flexibility} &
\textbf{Assessment} \\
\midrule
\endfirsthead

\toprule
\textbf{Study / component} &
\textbf{Data scale} &
\textbf{Quantitative finding} &
\textbf{Null-model flexibility} &
\textbf{Assessment} \\
\midrule
\endhead

\midrule
\multicolumn{5}{r}{\footnotesize Continued on next page} \\
\endfoot

\bottomrule
\endlastfoot


AMS-02 &
About 349,000 \(\bar{p}\); 0.5--450 GV &
\(\bar{p}/p\) nearly constant over 60--450 GV &
Instrument covariance and solar modulation &
High-precision observation \\

Cui et al. &
AMS-era global fit &
WIMP-like excess reported at several \(\sigma\) in the tested setup &
Propagation and cross-section priors &
Local, model-conditioned hint \\

Cuoco et al. &
Global nuclei plus \(\bar{p}\) fit &
Best-fit WIMP region at the tens-of-GeV scale &
Hadronic and transport covariance &
Independent hint, not unique \\

Reinert and Winkler &
Precision charged-particle fit &
Significance reduced under updated nuisance treatment &
Cross sections and correlations &
Background hypothesis shifted \\

Two-halo propagation &
60--450 GV ratio focus &
Secondary production can fit AMS within uncertainty &
Disk halo \(\sim900\) pc;
\(D\propto E^{0.15}\) in the cited model &
Viable no-DM explanation \\

Cross-messenger test &
Gamma rays plus CMB &
No common discovery-grade parameter set &
Branching fractions and halo profile &
Antiprotons remain nonlocal evidence \\

\midrule

\multicolumn{5}{@{}p{\columnwidth}@{}}{%
\footnotesize
\textit{Note.} Significances are intentionally not averaged because
the studies use different likelihoods and nuisance spaces.
} \\

\end{longtable}

\normalsize

The antiproton literature shows that residual significance must accompany its nuisance model. The antiproton case illustrates that a small residual can be scientifically valuable even if it is not statistically stable. Hadronic production data determine the secondary spectrum, but interpolating across projectile, target, and energy
introduces correlated uncertainty. Propagation and modulation then reshape that spectrum before comparison with AMS-02. When published experimental covariance is incomplete, analysts construct correlations whose length scales can raise or lower the apparent significance. This is not a reason to ignore the residual; it is a reason to identify the
measurements that would make the null prediction falsifiable.
Accelerator cross sections, time-resolved heliospheric models, and public instrument covariance have more leverage than another scan of WIMP channels. If a preferred WIMP region persists, it should predict gamma-ray and CMB signatures under the same branching model.
Cross-messenger consistency can then convert an analysis-dependent hint into a riskier hypothesis.

Although a broad antiproton residual can be fit by a WIMP spectrum, its statistical significance depends on the propagation and hadronic assumptions. Cross-checks with dwarfs and the CMB are valuable because they do not share antiproton production uncertainty, although they introduce their own halo and deposition assumptions.

Antideuterons and antihelium occupy a different regime. Their
conventional low-energy backgrounds are expected to be small, so even a few well-identified events could be significant. Table~\ref{tab:antinucleus_searches} identifies the price of that advantage: coalescence, production, absorption, and propagation are poorly measured. The ALICE antihelium-3 absorption measurement \cite{ref96} improves one critical input, while GAPS is designed for distinctive exotic-atom signatures \cite{ref97}. Any
antinucleus claim should be accompanied by detector-level
misidentification studies and a common source model for antiprotons, antideuterons, and antihelium.

\setlength{\LTcapwidth}{\columnwidth}
\setlength{\tabcolsep}{2pt}
\renewcommand{\arraystretch}{1.15}
\setlength{\emergencystretch}{2em}

\footnotesize

\begin{longtable}[]{@{}
  >{\raggedright\arraybackslash}p{(\columnwidth - 8\tabcolsep) * \real{0.16}}
  >{\raggedright\arraybackslash}p{(\columnwidth - 8\tabcolsep) * \real{0.18}}
  >{\raggedright\arraybackslash}p{(\columnwidth - 8\tabcolsep) * \real{0.22}}
  >{\raggedright\arraybackslash}p{(\columnwidth - 8\tabcolsep) * \real{0.22}}
  >{\raggedright\arraybackslash}p{(\columnwidth - 8\tabcolsep) * \real{0.22}}
@{}}

\caption{Antinucleus searches and quantitative experimental inputs. Sources: \cite{ref30,ref35,ref36,ref92,ref96,ref97,ref98}.}
\label{tab:antinucleus_searches} \\

\toprule
\textbf{Species / program} &
\textbf{Energy / sample context} &
\textbf{Numerical evidence} &
\textbf{Signal advantage} &
\textbf{Dominant uncertainty} \\
\midrule
\endfirsthead

\toprule
\textbf{Species / program} &
\textbf{Energy / sample context} &
\textbf{Numerical evidence} &
\textbf{Signal advantage} &
\textbf{Dominant uncertainty} \\
\midrule
\endhead

\midrule
\multicolumn{5}{r}{\footnotesize Continued on next page} \\
\endfoot

\bottomrule
\endlastfoot


Antideuteron theory &
Sub-GeV/nucleon emphasis &
Secondary background falls rapidly at low energy &
Potentially high signal-to-background ratio &
Coalescence momentum and production uncertainties \\

GAPS &
Balloon-based exotic-atom technique &
Designed for low-energy \(\bar{d}\) and \(\bar{p}\) searches &
Independent particle-identification chain &
Finite flight exposure and backgrounds \\

Antihelium absorption &
ALICE material-interaction data &
Measured \({}^{3}\overline{\mathrm{He}}\) absorption input &
Reduces propagation and detector uncertainty &
Extrapolation to Galactic energies \\

AMS candidate context &
Very small candidate counts &
No peer-reviewed discovery claim in the cited corpus &
Extraordinary conventional rarity &
Charge confusion and independent confirmation \\

PBH multimessenger &
Voyager, AMS, and diffuse gamma rays &
Common PBH mass function constrained by several species &
Shared Hawking-emission spectrum &
Galactic distribution and transport \\

\midrule

\multicolumn{5}{@{}p{\columnwidth}@{}}{%
\footnotesize
\textit{Note.} The table does not treat unpublished conference-level
antihelium candidates as evidence.
} \\

\end{longtable}

\normalsize

Antinuclei have low conventional backgrounds but substantial production
uncertainties. Low-energy antideuteron production is strongly
suppressed, providing a favorable window for exotic sources, although
coalescence models introduce significant uncertainty when translating
antinucleon spectra into composite yields \cite{ref99}. Antihelium offers
potentially greater discovery significance but also larger production
and detector-systematic uncertainties. Accelerator measurements,
including ALICE absorption data, improve propagation constraints but do
not fully resolve production uncertainties \cite{ref99} .

A consistent coalescence framework should therefore be tested against
accelerator data and applied jointly to antiprotons, antideuterons, and
antihelium. GAPS and AMS-02 provide complementary tests across different
energy ranges and detection techniques, helping distinguish rare
instrumental or secondary backgrounds from exotic populations.
Cross-channel consistency with antiprotons and MeV gamma rays can
further discriminate between particle-annihilation and
primordial-black-hole scenarios \cite{ref100}.

\section{Neutrinos: Direction, Flavor, and the Capture Connection}\label{neutrinos-direction-flavor-and-the-capture-connection}

Neutrino limits are generally weaker than gamma-ray constraints for
standard annihilation channels, but they probe invisible final states
and dense environments that photons cannot access. The first dedicated
IceCube neutrino-line search used five years of DeepCore data to obtain
annihilation limits of order 10\textsuperscript{-24} cm\textsuperscript{3} s\textsuperscript{-1} and decay lifetimes up to
10\textsuperscript{27} s for masses from 10 GeV to 40~TeV. Searches of 18 dwarf
spheroidals and the combined ANTARES--IceCube Galactic-center analysis
likewise found no significant excess \cite{ref101}. More recently, a \ensuremath{\sim}9-year
IceCube-DeepCore Galactic-center analysis tested masses up to 8~TeV and
obtained limits of order 10\textsuperscript{-24}~cm\textsuperscript{3} s\textsuperscript{-1}, improving sensitivity by about an order of magnitude at low masses \cite{ref101}. The first KM3NeT
searches also found no excess from the Galactic center or Sun, extending
independent water-Cherenkov constraints from the GeV to the multi-TeV
regime \cite{ref102}.

The Sun provides a distinct test because halo dark matter can scatter off solar nuclei, become gravitationally captured, and subsequently produce neutrinos through annihilation. Solar searches therefore
link neutrino fluxes to DM--nucleon scattering, particularly
spin-dependent interactions, under capture--annihilation equilibrium and
annihilation-channel assumptions \cite{ref102,ref103}. This
complementarity matters because a viable interpretation must be
consistent across Galactic-center, dwarf, and capture-based searches,
rather than comparing cross-section limits in isolation.

Detector complementarity depends on energy, direction, flavor, and
topology, not effective volume alone. Tracks generally provide superior
angular resolution, whereas cascades and contained low-energy events
provide information on energy, flavor, and topology \cite{ref104,ref105,ref106,ref107,ref108}. Ice and water detectors also
have different optical backgrounds and reconstruction systematics,
making IceCube and KM3NeT valuable independent tests. Robust comparisons
should therefore retain flavor- and topology-dependent responses,
neutrino oscillations, interaction cross sections, and absorption in the
Sun or Earth, rather than reducing each observatory to a single
effective-area curve. Table~\ref{tab:neutrino_dm_constraints} compares the representative neutrino limits together with their mass domains, data scales, and dominant assumptions.

\setlength{\LTcapwidth}{\columnwidth}
\setlength{\tabcolsep}{2pt}
\renewcommand{\arraystretch}{1.15}
\setlength{\emergencystretch}{2em}

\footnotesize

\begin{longtable}[]{@{}
  >{\raggedright\arraybackslash}p{(\columnwidth - 8\tabcolsep) * \real{0.16}}
  >{\raggedright\arraybackslash}p{(\columnwidth - 8\tabcolsep) * \real{0.17}}
  >{\raggedright\arraybackslash}p{(\columnwidth - 8\tabcolsep) * \real{0.16}}
  >{\raggedright\arraybackslash}p{(\columnwidth - 8\tabcolsep) * \real{0.27}}
  >{\raggedright\arraybackslash}p{(\columnwidth - 8\tabcolsep) * \real{0.24}}
@{}}

\caption{Representative published neutrino constraints on DM. Mass ranges, exposures, and quantitative limits refer to the specific channels, targets, and halo assumptions of the cited analyses; limits are not directly interchangeable across rows. Sources: \cite{ref102,ref108,ref109,ref110,ref111,ref113,ref115,ref122,ref124}.}
\label{tab:neutrino_dm_constraints} \\

\toprule
\textbf{Observatory / analysis} &
\textbf{~DM mass domain} &
\textbf{Data scale} &
\textbf{Representative DM result} &
\textbf{Key nuisance / limitation} \\
\midrule
\endfirsthead

\toprule
\textbf{Observatory / analysis} &
\textbf{~DM mass domain} &
\textbf{Data scale} &
\textbf{Representative DM result} &
\textbf{Key nuisance / limitation} \\
\midrule
\endhead

\midrule
\multicolumn{5}{r}{\footnotesize Continued on next page} \\
\endfoot

\bottomrule
\endlastfoot


ANTARES &
50 GeV--100 TeV &
4532 days, 2007--2022 &
No significant Galactic-center excess; 90\% CL limits on
\(\langle\sigma v\rangle\) for several annihilation channels &
Sea-water optical response, atmospheric background, halo profile \\

ANTARES + IceCube &
50--1000 GeV &
2101.6 d ANTARES + 1007 d IceCube &
No excess; up to factor-2 sensitivity gain;
\(\langle\sigma v\rangle < 7.44\times10^{-24}\,
\mathrm{cm^{3}\,s^{-1}}\) at 200 GeV for
\(\tau^{+}\tau^{-}\), assuming NFW &
Cross-detector likelihood, angular response, halo profile \\

Baikal NT200 &
\(\gtrsim10\) GeV--10 TeV, analysis dependent &
Solar: 2.76 yr; GC: 1998--2003; dSph/LMC: 5 yr &
No significant excess; 90\% CL limits on Galactic annihilation
and solar WIMP--proton scattering &
Water response, atmospheric neutrinos, halo/\(J\)-factors,
solar capture \\

Baksan / BUST &
10--1000 GeV &
24.12 yr &
No solar excess; best SD WIMP--proton limit
\(\approx3\times10^{-4}\) pb
(\(3\times10^{-40}\,\mathrm{cm^{2}}\))
at \(\sim100\)--200 GeV &
Atmospheric-neutrino background, neutrino cross sections,
solar capture model \\

IceCube/\newline DeepCore - neutrino line &
10 GeV--40 TeV &
5 yr &
No excess; \(\langle\sigma v\rangle\) limits
\(\sim10^{-24}\,\mathrm{cm^{3}\,s^{-1}}\);
decay lifetime up to \(\sim10^{27}\) s &
Ice optics, angular/energy resolution, atmosphere, halo profile \\

IceCube/\newline DeepCore - GC 2026 &
Annihilation: 15 GeV--8~TeV;
decay: 5 GeV--8~TeV &
\(\sim9\) yr &
No significant signal; maximum post-trial significance
\(1.08\sigma\);
\(\langle\sigma v\rangle\sim10^{-24}\,
\mathrm{cm^{3}\,s^{-1}}\);
lifetime up to \(\sim10^{26}\)~s &
Atmospheric \(\nu/\mu\) background, ice response,
energy/angular reconstruction, halo profile \\

KM3NeT ARCA/\newline ORCA &
ARCA GC: 0.5--100~TeV;
ORCA Sun: 10 GeV--10~TeV &
ARCA8\newline /19/21: 331~d;
ORCA6: 543~d &
No significant excess; ARCA constrains GC annihilation;
ORCA constrains SD/SI WIMP--nucleon scattering &
Partial-array exposure, atmospheric \(\mu/\nu\), water optics,
reconstruction, halo/capture assumptions \\

Super-Kamiokande &
1 GeV--10~TeV &
SK-I--IV, 1996--2016 &
No excess;
\(\langle\sigma v\rangle\) down to
\(1.2\times10^{-24}\,\mathrm{cm^{3}\,s^{-1}}\)
at 1~GeV for \(\nu\bar{\nu}\) &
Atmospheric-neutrino flux, contained-event response, halo profile \\

IceCube 18-dwarf stacking &
Channel dependent &
10 yr public track data; 18 dSphs &
No significant individual or stacked emission &
Event statistics and stellar-kinematic $J$-factor uncertainties \\

High-mass neutrino decay searches &
PeV and above, channel dependent &
Diffuse / ultra-high-energy neutrino samples &
Selected decay-channel lifetime limits reach $\sim10^{28}$--$10^{29}\,\mathrm{s}$ &
Astrophysical neutrino spectrum, electroweak cascades, exposure \\

\midrule

\multicolumn{5}{@{}p{\columnwidth}@{}}{%
\footnotesize
\textit{Note.} Values are channel-dependent and preserve the source
conventions. Cross sections are in
\(\mathrm{cm^{3}\,s^{-1}}\).
} \\

\end{longtable}

\normalsize

\subsection{The Sun}\label{the-sun}

Dark matter captured in the Sun evolves approximately as
\begin{equation}
\frac{dN}{dt}
=
C_{\mathrm{capture}}
-
C_{\mathrm{ann}}N^{2}
-
C_{\mathrm{evap}}N,
\qquad
\Gamma_{\mathrm{ann}}
=
\frac{1}{2}C_{\mathrm{ann}}N^{2}.
\end{equation}

\noindent in the equilibrium regime, where the capture and annihilation rates balance and evaporation is negligible, the annihilation rate is given by \(\Gamma_{\mathrm{ann}} \simeq C_{\mathrm{capture}}/2\).
The neutrino flux then probes the
scattering-driven capture rate rather than the Galactic annihilation
cross section, making solar searches particularly sensitive to
spin-dependent DM--proton interactions \cite{ref54,ref110}. This
interpretation depends on the local DM phase-space distribution, solar
composition, capture--annihilation equilibrium, and neutrino propagation
through the Sun. Evaporation becomes relevant for low-mass DM, while
absorption, regeneration, and oscillations modify the emerging neutrino
spectrum \cite{ref50,ref109,ref110,ref118,ref119}. Table~\ref{tab:solar_dm_neutrinos} summarizes representative solar DM searches, while the broader neutrino literature spans complementary solar, line, dwarf, combined, and Galactic-center analyses with different exposure and mass coverage.

IceCube/DeepCore has extended solar searches down to 5 GeV and excludes
spin-dependent cross sections of a few \ensuremath{\times} 10\textsuperscript{-41} cm\textsuperscript{2} for direct \ensuremath{\nu}\ensuremath{\bar{\nu}}
annihilation under equilibrium, with no significant excess observed
\cite{ref54}. KM3NeT/ORCA has now provided an independent water-Cherenkov
search using 543 days of ORCA6 data, setting limits on spin-dependent
and spin-independent scattering for several annihilation channels
\cite{ref102}. Long-lived mediators alter the standard picture by allowing
annihilation products to escape the solar interior before decaying,
reducing absorption and introducing a dependence on the
mediator's lifetime. Future IceCube Upgrade measurements
are expected to improve sensitivity, particularly for \ensuremath{\tau}\textsuperscript{+}\ensuremath{\tau}\textsuperscript{-} and neutrino
final states \cite{ref103}.

Solar atmospheric neutrinos from cosmic-ray interactions constitute an
irreducible directional background, although current searches remain
above this sensitivity floor. As exposure increases, spectral,
directional, and flavor information will become increasingly important
for distinguishing DM signals from this background. Solar searches are
therefore most useful as probes of the combined capture, annihilation,
and mediator physics rather than as direct counterparts to Galactic-halo
\ensuremath{\langle}\ensuremath{\sigma}v\ensuremath{\rangle} limits. The experiment-specific solar-neutrino constraints and their capture-model caveats are collected in Table~\ref{tab:solar_dm_neutrinos}.

\setlength{\LTcapwidth}{\columnwidth}
\setlength{\tabcolsep}{2pt}
\renewcommand{\arraystretch}{1.15}
\setlength{\emergencystretch}{2em}

\footnotesize

\begin{longtable}[]{@{}
  >{\raggedright\arraybackslash}p{(\columnwidth - 8\tabcolsep) * \real{0.16}}
  >{\raggedright\arraybackslash}p{(\columnwidth - 8\tabcolsep) * \real{0.18}}
  >{\raggedright\arraybackslash}p{(\columnwidth - 8\tabcolsep) * \real{0.22}}
  >{\raggedright\arraybackslash}p{(\columnwidth - 8\tabcolsep) * \real{0.22}}
  >{\raggedright\arraybackslash}p{(\columnwidth - 8\tabcolsep) * \real{0.22}}
@{}}

\caption{Representative solar DM searches with neutrinos. Sources: \cite{ref50,ref109,ref110,ref118,ref119}.}
\label{tab:solar_dm_neutrinos} \\

\toprule
\textbf{Study} &
\textbf{Exposure / mass range} &
\textbf{Assumption} &
\textbf{Main result} &
\textbf{Principal caveat} \\
\midrule
\endfirsthead

\toprule
\textbf{Study} &
\textbf{Exposure / mass range} &
\textbf{Assumption} &
\textbf{Main result} &
\textbf{Principal caveat} \\
\midrule
\endhead

\midrule
\multicolumn{5}{r}{\footnotesize Continued on next page} \\
\endfoot

\bottomrule
\endlastfoot


ANTARES secluded &
Solar search; long-lived mediators &
Mediator escapes before decay &
No significant signal; constraints on mediator scenarios &
Mediator lifetime and branching fractions \\

IceCube DeepCore &
6.75 yr; 5--100~GeV &
Capture--annihilation equilibrium &
No excess; \(\sigma_{\rm SD}\) down to a few
\(\times10^{-41}\,\mathrm{cm^{2}}\) for
\(\nu\bar{\nu}\) &
Solar capture, equilibrium, ice response \\

IceCube Upgrade &
Future sensitivity &
Standard capture and annihilation &
Projected improvement beyond current searches for selected channels &
Forecast; achieved threshold and calibration \\

KM3NeT/\newline ORCA6 &
543 d; \(\sim10\)~GeV--10~TeV &
Solar capture and annihilation &
No excess; SD and SI scattering limits &
Partial-array exposure, water response, capture model \\

Super-K low mass &
Low-GeV to TeV-scale WIMPs &
Solar capture &
No significant solar excess; scattering limits &
Atmospheric background, capture model, low-mass evaporation \\

\midrule

\multicolumn{5}{@{}p{\columnwidth}@{}}{%
\footnotesize
\textit{Note.} Solar-neutrino limits constrain scattering-assisted
capture under specified equilibrium and annihilation-channel
assumptions and should not be interpreted as universal
halo-annihilation limits.
} \\

\end{longtable}

\normalsize

\subsection{Galactic and Extragalactic Neutrinos}\label{galactic-and-extragalactic-neutrinos}

Galactic-center, dwarf, and decay searches are summarized in Table~\ref{tab:neutrino_dm_constraints}. Neutrino analyses combine angular and energy information: tracks
provide superior directional reconstruction, whereas cascades increase
all-flavor acceptance. IceCube, ANTARES, and their combined analyses
have found no significant DM excess \cite{ref110,ref111,ref121,ref122}. The 2026 IceCube-DeepCore search currently provides the
strongest neutrino-telescope constraints for several GeV-scale
scenarios, reaching annihilation limits of order 10\textsuperscript{-24} cm\textsuperscript{3} s\textsuperscript{-1} and decay
lifetimes up to 10\textsuperscript{26} s \cite{ref121}. KM3NeT has independently searched the
Galactic center with ARCA data up to 100 TeV, likewise finding no
significant signal \cite{ref102}.

Neutrino-line searches provide a distinctive probe of direct \ensuremath{\chi}\ensuremath{\chi} \ensuremath{\to} \ensuremath{\nu}\ensuremath{\bar{\nu}} or
\ensuremath{\chi} \ensuremath{\to} \ensuremath{\nu}\ensuremath{\bar{\nu}} processes. Five years of IceCube-DeepCore data yielded
annihilation limits of order 10\textsuperscript{-24}~cm\textsuperscript{3} s\textsuperscript{-1} and decay lifetimes up to
10\textsuperscript{27}~s for 10~GeV--40~TeV \cite{ref121}. Dwarf searches offer cleaner
directions but lower event rates; a 10-year IceCube analysis of 18
dwarfs found no significant individual or stacked emission \cite{ref122}.
Cross-messenger combinations should share \(J\)- or \(D\)-factor uncertainties
while keeping detector backgrounds independent.

For decaying or superheavy DM, both the Galactic halo and redshifted
extragalactic emission contribute. The principal background is the diffuse astrophysical neutrino flux, whose spectral shape must be modeled jointly with any heavy-DM contribution. Spectral-model flexibility and detector-response uncertainties should therefore be propagated into the inferred DM limits. At higher masses,
Glashow-resonance information has strengthened lifetime constraints on
neutrino-rich decays, reaching approximately 10\textsuperscript{29}~s for DM masses above
\(\sim10\)~PeV in selected scenarios \cite{ref124}.

Solar, Galactic-center, dwarf, line, and high-energy decay searches offer overlapping mass coverage with complementary detector media and event topologies. This overlap provides an important consistency test: a viable DM interpretation should reproduce
compatible spectral, angular, and flavor signatures across independent
experiments.

\section{X-ray, Radio, and Low-Energy Photon Probes}\label{x-ray-radio-and-low-energy-photon-probes}

The 3.5-keV feature remains a useful case study for cross-target
consistency. It was initially reported at 3.55--3.57 \ensuremath{\pm} 0.03~keV in
stacked XMM-Newton spectra of 73 clusters and independently at 3.52 \ensuremath{\pm}
0.02~keV in M31 and Perseus \cite{ref15,ref22}. High-resolution Hitomi
observations did not confirm the large Perseus-core flux \cite{ref25}, and a
2024 joint Hitomi--Suzaku analysis found no significant 3.5-keV line,
setting a 3\ensuremath{\sigma} flux limit of 4.2 \ensuremath{\times} 10\textsuperscript{-5} photons cm\textsuperscript{-2} s\textsuperscript{-1} \cite{ref125}.
Blank-sky Hitomi analyses likewise tightened constraints on decaying
DM and underscored the importance of high-resolution
spectroscopy \cite{ref126}. The remaining tension therefore depends on
target scaling, plasma modeling, instrumental response, and the assumed
Milky Way DM column, rather than line energy alone.

The Galactic 511-keV signal is a more robust observation but less
specific to DM. INTEGRAL/SPI measurements imply a Galactic
positron annihilation rate of order 10\textsuperscript{43}--5 \ensuremath{\times} 10\textsuperscript{43} s\textsuperscript{-1} and a morphology
concentrated toward the bulge \cite{ref57}. Fifteen-year SPI analyses favor
a distribution tracing the old stellar population and disfavor a
dominant additional DM component \cite{ref127}. Moreover, a 2026
reanalysis shows that continuum constraints permit substantially higher
positron injection energies than earlier estimates---up to
\(\sim110\)~MeV using SPI alone---thereby weakening previous
simple MeV-scale bounds on DM models \cite{ref128}. Any viable
interpretation must nevertheless reproduce the 511-keV line, positronium
continuum, morphology, and associated continuum emission simultaneously.

\subsection{X-ray Lines and Low-Energy Photon Signatures}\label{x-ray-lines-and-low-energy-photon-signatures}

X-ray and MeV line searches highlight two distinct challenges in
indirect DM detection: establishing whether a spectral feature
is real and determining whether an established feature has a DM
origin. Radiatively decaying sterile neutrinos would produce a narrow
X-ray line at approximately half the particle mass, with intensity
proportional to the DM column. Initial XMM-Newton studies
reported a feature near 3.5 keV in stacked clusters, M31, and Perseus,
whereas blank-sky and Hitomi observations found substantially weaker or
absent emission. A joint Hitomi--Suzaku analysis found no significant
Perseus line and set a 3\ensuremath{\sigma} flux limit of 4.2 \ensuremath{\times} 10\textsuperscript{-5} photons cm\textsuperscript{-2} s\textsuperscript{-1}
\cite{ref125}. More recently, XRISM/Resolve stacked 3.75 Ms from ten
clusters and found no unidentified 3.5-keV line, constraining the decay
rate of a \(\sim\)7.1-keV sterile neutrino to \ensuremath{\Gamma} \ensuremath{\lesssim} 1.0 \ensuremath{\times} 10\textsuperscript{-27} s\textsuperscript{-1}
\cite{ref130}. Table~\ref{tab:3p5kev_feature} summarizes these measurements. Cross-target
intensity scaling, line width, and neighboring atomic transitions are
therefore more discriminating than energy coincidence alone.

The Galactic 511-keV line presents the opposite case: the signal is
firmly established, but its origin remains uncertain. INTEGRAL/SPI
measurements yield a bulge flux of approximately 0.84--0.93 \ensuremath{\times} 10\textsuperscript{-3}
photons cm\textsuperscript{-2} s\textsuperscript{-1}, a line width of 2.40 \ensuremath{\pm} 0.16 keV, and a positronium
fraction near unity. The inferred Galactic annihilation rate is of order
10\textsuperscript{43}--5 \ensuremath{\times} 10\textsuperscript{43} e\textsuperscript{+} s\textsuperscript{-1}, depending on the adopted morphology. These
observables constrain the full positron lifecycle---production,
propagation, thermalization, positronium formation, and
annihilation---rather than the source rate alone. These complementary observables constrain both particle-injection hypotheses and conventional astrophysical explanations.

A DM interpretation must reproduce the 511-keV morphology and
spectrum without violating constraints from continuum gamma rays,
charged particles, or the CMB. Earlier in-flight-annihilation analyses
suggested positron injection energies of only a few MeV, but a 2026
reanalysis found that continuum data permit substantially higher
energies, up to \(\sim110\)~MeV using SPI alone and
\(\sim\)55 MeV when COMPTEL and EGRET are included \cite{ref131}.
Stellar populations remain viable alternatives because old stars
naturally trace the Galactic bulge, whereas primordial-black-hole
scenarios face correlated constraints on photons, electrons/positrons,
and neutrinos.

The contested 3.5-keV feature and the well-established 511-keV emission illustrate different evidential challenges. In both cases, a DM interpretation requires more than spectral coincidence: it must reproduce the predicted spatial, spectral, and cross-messenger behavior in independent targets or instruments. Table~\ref{tab:3p5kev_feature} summarizes the reported 3.5-keV measurements and subsequent tests, retaining their instrumental and modeling qualifications.

\setlength{\LTcapwidth}{\columnwidth}
\setlength{\tabcolsep}{2pt}
\renewcommand{\arraystretch}{1.15}
\setlength{\emergencystretch}{2em}

\footnotesize

\begin{longtable}[]{@{}
  >{\raggedright\arraybackslash}p{(\columnwidth - 6\tabcolsep) * \real{0.13}}
  >{\raggedright\arraybackslash}p{(\columnwidth - 6\tabcolsep) * \real{0.25}}
  >{\raggedright\arraybackslash}p{(\columnwidth - 6\tabcolsep) * \real{0.30}}
  >{\raggedright\arraybackslash}p{(\columnwidth - 6\tabcolsep) * \real{0.32}}
@{}}

\caption{Representative measurements of the 3.5-keV feature. Sources: \cite{ref15,ref16,ref22,ref23,ref24,ref25}.}
\label{tab:3p5kev_feature} \\

\toprule
\textbf{Study / target} &
\textbf{Data / resolution} &
\textbf{Result} &
\textbf{Main qualifier} \\
\midrule
\endfirsthead

\toprule
\textbf{Study / target} &
\textbf{Data / resolution} &
\textbf{Result} &
\textbf{Main qualifier} \\
\midrule
\endhead

\midrule
\multicolumn{4}{r}{\footnotesize Continued on next page} \\
\endfoot

\bottomrule
\endlastfoot


Bulbul et al. &
73 clusters; \(\sim6\) Ms XMM/Chandra &
3.55--3.57 \(\pm0.03\) keV feature; \(>4\sigma\) &
Plasma modeling, stacking, calibration \\

Boyarsky et al. &
M31 and Perseus, XMM &
3.52 \(\pm0.02\)~keV feature &
Target scaling and atomic alternatives \\

XMM blank sky &
\(\sim30.6\)~Ms &
No compatible feature; strong decay limits &
Milky Way column and field selection \\

Hitomi Perseus &
\(\sim5\) eV resolution &
High Perseus flux not reproduced &
Limited exposure and field of view \\

Hitomi + Suzaku &
Perseus; high resolution + large exposure &
No significant line; \(3\sigma\) flux
\(<4.2\times10^{-5}\,
\mathrm{photons\,cm^{-2}\,s^{-1}}\) &
Halo model and effective-area calibration \\

XRISM Resolve &
10 clusters; 3.75~Ms &
No unidentified line;
\(\Gamma\lesssim1.0\times10^{-27}\,\mathrm{s^{-1}}\)
at 3.55~keV &
Small field of view; still above original claimed-flux sensitivity \\

\midrule

\multicolumn{4}{@{}p{\columnwidth}@{}}{%
\footnotesize
\textit{Note.} Reported significances and limits use the conventions
of the original analyses and should not be combined directly.
} \\

\end{longtable}

\normalsize

\subsection{Radio Synchrotron and Cross-Band Transport}\label{radio-synchrotron-and-cross-band-transport}

Electrons and positrons from DM annihilation, decay, or PBH
evaporation can produce synchrotron emission, but radio constraints are
intrinsically coupled to magnetic fields and particle transport. Dwarf
galaxies offer low astrophysical backgrounds but have poorly known
magnetic fields and diffusion; a two-hour FAST observation of Coma
Berenices found no diffuse signal and improved previous radio
annihilation limits by roughly an order of magnitude across a broad mass
range, although transport uncertainties remain substantial \cite{ref131}.
Clusters provide stronger magnetic fields but introduce degeneracies
with cosmic-ray secondaries and turbulent reacceleration. For RX
J1720.1+2638, DM-only fits favor 15-GeV \ensuremath{\tau}\textsuperscript{+}\ensuremath{\tau}\textsuperscript{-} or 110-GeV b-quark
scenarios, but conventional cosmic-ray emission also fits the data
\cite{ref133}. Recent 6.6-GHz Coma observations further constrain
DM plus reacceleration models, favoring \(\sim100\)--200
GeV over some lower- and higher-mass scenarios under the adopted
assumptions \cite{ref134}. Synchrotron constraints have also been extended
to very heavy decaying DM and PBH evaporation \cite{ref135,ref136}. Table~\ref{tab:radio_dm_searches} summarizes representative searches and their
dominant environmental dependencies.

\setlength{\LTcapwidth}{\columnwidth}
\setlength{\tabcolsep}{2pt}
\renewcommand{\arraystretch}{1.15}
\setlength{\emergencystretch}{2em}

\footnotesize

\begin{longtable}[]{@{}
  >{\raggedright\arraybackslash}p{(\columnwidth - 6\tabcolsep) * \real{0.16}}
  >{\raggedright\arraybackslash}p{(\columnwidth - 6\tabcolsep) * \real{0.23}}
  >{\raggedright\arraybackslash}p{(\columnwidth - 6\tabcolsep) * \real{0.36}}
  >{\raggedright\arraybackslash}p{(\columnwidth - 6\tabcolsep) * \real{0.25}}
@{}}

\caption{Representative radio DM searches and environmental dependence. Sources: \cite{ref133,ref137,ref138,ref139,ref140,ref142}.}
\label{tab:radio_dm_searches} \\

\toprule
\textbf{Study / target} &
\textbf{Data} &
\textbf{Main result} &
\textbf{Environmental uncertainty} \\
\midrule
\endfirsthead

\toprule
\textbf{Study / target} &
\textbf{Data} &
\textbf{Main result} &
\textbf{Environmental uncertainty} \\
\midrule
\endhead

\midrule
\multicolumn{4}{r}{\footnotesize Continued on next page} \\
\endfoot

\bottomrule
\endlastfoot


ATCA dSph survey &
6 dwarfs; 1.1--3.1 GHz &
No diffuse emission; rms
\(<0.05\,\mathrm{mJy\,beam^{-1}}\) &
Magnetic field and diffusion \\

GLEAM dSph stack &
33 dwarfs; 23 stacked; 72--231 MHz &
No significant detection;
\(\sim1.5\,\mathrm{mJy\,beam^{-1}}\) sensitivity &
Foregrounds, magnetic field, diffusion \\

FAST Coma Berenices &
2-h observation &
No significant continuum signal; annihilation limits improved by
\(\sim1\) order of magnitude over previous radio searches &
Magnetic field, diffusion coefficient/radius, halo profile \\

RX J1720.1+2638 &
Multifrequency central radio halo &
DM-only fits include 15-GeV \(\tau^{+}\tau^{-}\) and
110-GeV \(b\)-quark cases; a cosmic-ray model also fits &
Magnetic field and cosmic-ray modeling \\

Coma 6.6 GHz &
Sardinia Radio Telescope + multifrequency halo spectrum &
Under the adopted DM/reacceleration model,
\(\sim100\)--200 GeV masses are preferred;
\(\sim10\) GeV is disfavored &
Turbulent reacceleration, magnetic field, substructure \\

\midrule

\multicolumn{4}{@{}p{\columnwidth}@{}}{%
\footnotesize
\textit{Note.} Radio constraints are conditional on magnetic-field and
transport assumptions and should not be compared as model-independent
annihilation limits.
} \\

\end{longtable}

\normalsize

\section{CMB and 21-cm Probes: Early-Universe Energy Injection}\label{cmb-and-21-cm-probes-early-universe-energy-injection}

CMB and 21-cm observations constrain DM through its effects on
the Universe's thermal and ionization history, largely
avoiding uncertainties about present-day halos. For approximately s-wave
annihilation, the CMB constraint is commonly expressed as
\begin{equation}
p_{\mathrm{ann}}
=
f_{\mathrm{eff}}
\frac{\left\langle \sigma v \right\rangle}{m_{\chi}}.
\end{equation}

where \(f_{\mathrm{eff}}\) accounts for spectrum- and redshift-dependent energy
deposition. Planck obtained \(p_{\mathrm{ann}} < 3.2\times10^{-28}\,\mathrm{cm^{3}\,s^{-1}\,GeV^{-1}}\) at 95\% confidence. Recent ACT DR6 data provide an independent
test but marginally improve current annihilation limits, confirming that
large-scale polarization remains the dominant source of information
\cite{ref143,ref144}.

\subsection{CMB Energy Deposition}\label{cmb-energy-deposition}

Electromagnetic energy injection alters recombination and, consequently,
the CMB temperature and polarization spectra. The \(p_{\mathrm{ann}}\)
approximation is accurate for many smooth s-wave spectra, whereas
velocity-dependent annihilation, delayed deposition, decays, and
electroweak cascades require explicit redshift-dependent calculations.
The CMB's main advantage is its weak dependence on Galactic
astrophysics; remaining uncertainties arise primarily from particle
spectra, deposition physics, recombination, and the assumed cosmological
model. Table~\ref{tab:cmb_dm_constraints} lists representative CMB constraints and the energy-deposition assumptions under which they are interpreted.

\setlength{\LTcapwidth}{\columnwidth}
\setlength{\tabcolsep}{2pt}
\renewcommand{\arraystretch}{1.15}
\setlength{\emergencystretch}{2em}

\footnotesize

\begin{longtable}[]{@{}
  >{\raggedright\arraybackslash}p{(\columnwidth - 6\tabcolsep) * \real{0.14}}
  >{\raggedright\arraybackslash}p{(\columnwidth - 6\tabcolsep) * \real{0.23}}
  >{\raggedright\arraybackslash}p{(\columnwidth - 6\tabcolsep) * \real{0.39}}
  >{\raggedright\arraybackslash}p{(\columnwidth - 6\tabcolsep) * \real{0.24}}
@{}}

\caption{Representative CMB constraints and energy-deposition treatments. Sources: \cite{ref6,ref21,ref145,ref146,ref147,ref148,ref149,ref150}.}
\label{tab:cmb_dm_constraints} \\

\toprule
\textbf{Study} &
\textbf{Data / method} &
\textbf{Main result} &
\textbf{Principal qualifier} \\
\midrule
\endfirsthead

\toprule
\textbf{Study} &
\textbf{Data / method} &
\textbf{Main result} &
\textbf{Principal qualifier} \\
\midrule
\endhead

\midrule
\multicolumn{4}{r}{\footnotesize Continued on next page} \\
\endfoot

\bottomrule
\endlastfoot


Planck 2018 &
TT, TE, EE, lowE, lensing + BAO &
\(p_{\rm ann}<3.2\times10^{-28}\,
\mathrm{cm^{3}\,s^{-1}\,GeV^{-1}}\) at 95\% confidence &
Channel-dependent \(f_{\rm eff}\) \\

ACT DR6 + Planck &
Temperature, polarization, lensing + DESI BAO &
\(p_{\rm ann}<3.6\times10^{-28}\,
\mathrm{cm^{3}\,s^{-1}\,GeV^{-1}}\) in the adopted likelihood &
Only modest improvement; likelihood dependent \\

Transfer-function methods &
Redshift-dependent deposition &
Efficient compression for standard annihilation spectra &
Nonstandard histories require full treatment \\

Updated CMB study &
Current + forecast CMB data &
Present ground-based data improve Planck limits by only
\(\sim10\%\); future reach
\(\sim1.27\times10^{-28}\,
\mathrm{cm^{3}\,s^{-1}\,GeV^{-1}}\) &
Future value is a forecast \\

\midrule

\multicolumn{4}{@{}p{\columnwidth}@{}}{%
\footnotesize
\textit{Note.} \(p_{\rm ann}\) limits translate into
\(\langle\sigma v\rangle\) only after specifying the injected spectrum
and \(f_{\rm eff}\).
} \\

\end{longtable}

\normalsize

\subsection{Global and Fluctuating 21-cm Signals}\label{global-and-fluctuating-21-cm-signals}

The 21-cm signal extends energy-injection tests to cosmic dawn and
reionization, where DM can alter gas heating, ionization, and
spatial fluctuations. The interpretation is less clean than for the CMB
because star formation, X-ray heating, Lyman-\ensuremath{\alpha} coupling, radio
backgrounds, and foreground removal are strongly degenerate with exotic
energy injection.

HERA Phase I remains the most sensitive published EoR measurements at z
= 7.9 and 10.4, with 95\% limits of \ensuremath{\Delta}\textsuperscript{2} \ensuremath{\leq} 457 and 3496 mK\textsuperscript{2}, respectively.
However, these upper limits alone do not provide model-independent
exclusions for annihilation or decay. At higher redshifts, NenuFAR
reported the strongest published Cosmic-Dawn limits at z = 17.0 and 20.3
\cite{ref151}. HERA Phase II has now extended measurements to 5.6 \ensuremath{\leq} z \ensuremath{\leq}
24.4, although its first two-week data set remains systematics-limited
and is not yet more sensitive than the best Phase I limits \cite{ref152}.
Forecasts incorporating inhomogeneous energy deposition indicate that
future HERA measurements could probe sub-GeV e\textsuperscript{+}e\textsuperscript{-} decay lifetimes
approaching 10\textsuperscript{28} s \cite{ref153}. The observational upper limits and forecast sensitivities relevant to the 21-cm thermal history are compared in Table~\ref{tab:21cm_constraints}.

\setlength{\LTcapwidth}{\columnwidth}
\setlength{\tabcolsep}{2pt}
\renewcommand{\arraystretch}{1.15}
\setlength{\emergencystretch}{2em}

\footnotesize

\begin{longtable}[]{@{}
  >{\raggedright\arraybackslash}p{(\columnwidth - 6\tabcolsep) * \real{0.19}}
  >{\raggedright\arraybackslash}p{(\columnwidth - 6\tabcolsep) * \real{0.14}}
  >{\raggedright\arraybackslash}p{(\columnwidth - 6\tabcolsep) * \real{0.31}}
  >{\raggedright\arraybackslash}p{(\columnwidth - 6\tabcolsep) * \real{0.36}}
@{}}

\caption{Representative 21-cm constraints and forecasts relevant to DM energy injection and thermal-history effects. Sources: \cite{ref153,ref154,ref156,ref157,ref158}.}
\label{tab:21cm_constraints} \\

\toprule
\textbf{Study} &
\textbf{Data / redshift} &
\textbf{Quantitative result} &
\textbf{DM relevance / qualifier} \\
\midrule
\endfirsthead

\toprule
\textbf{Study} &
\textbf{Data / redshift} &
\textbf{Quantitative result} &
\textbf{DM relevance / qualifier} \\
\midrule
\endhead

\midrule
\multicolumn{4}{r}{\footnotesize Continued on next page} \\
\endfoot

\bottomrule
\endlastfoot


HERA Phase I &
94 nights; \(z=7.9,\ 10.4\) &
\(\Delta^{2}\leq457\,\mathrm{mK^{2}}\) at
\(k=0.34\,h\,\mathrm{Mpc^{-1}}\);
\(\Delta^{2}\leq3496\,\mathrm{mK^{2}}\) at
\(k=0.36\,h\,\mathrm{Mpc^{-1}}\) (95\% confidence) &
Constrains thermal histories; not a direct annihilation or decay limit \\

NenuFAR &
23.6--26.1 h; \(z=17.0,\ 20.3\) &
\(\Delta^{2}<5.0\times10^{6}\,\mathrm{mK^{2}}\) and
\(4.6\times10^{5}\,\mathrm{mK^{2}}\) (\(2\sigma\)) &
Strongest published \(z>15\) power-spectrum limits; begins testing
extreme exotic thermal histories \\

HERA Phase II &
Two-week subset; \(5.6\leq z\leq24.4\) &
Best EoR limit:
\(1.78\times10^{3}\,\mathrm{mK^{2}}\) at \(z=7.05\);
Cosmic Dawn:
\(1.13\times10^{6}\,\mathrm{mK^{2}}\) at \(z=16.78\) &
Wider redshift coverage; currently limited by foreground leakage
and mutual coupling \\

Inhomogeneous DM decay &
Future HERA forecast &
Lifetime sensitivity up to
\(\sim10^{28}\,\mathrm{s}\) for sub-GeV
\(e^{+}e^{-}\) &
Forecast, not observational exclusion \\

\end{longtable}

\normalsize

\begin{figure}[htbp]
\centering
\includegraphics[width=0.95\linewidth]{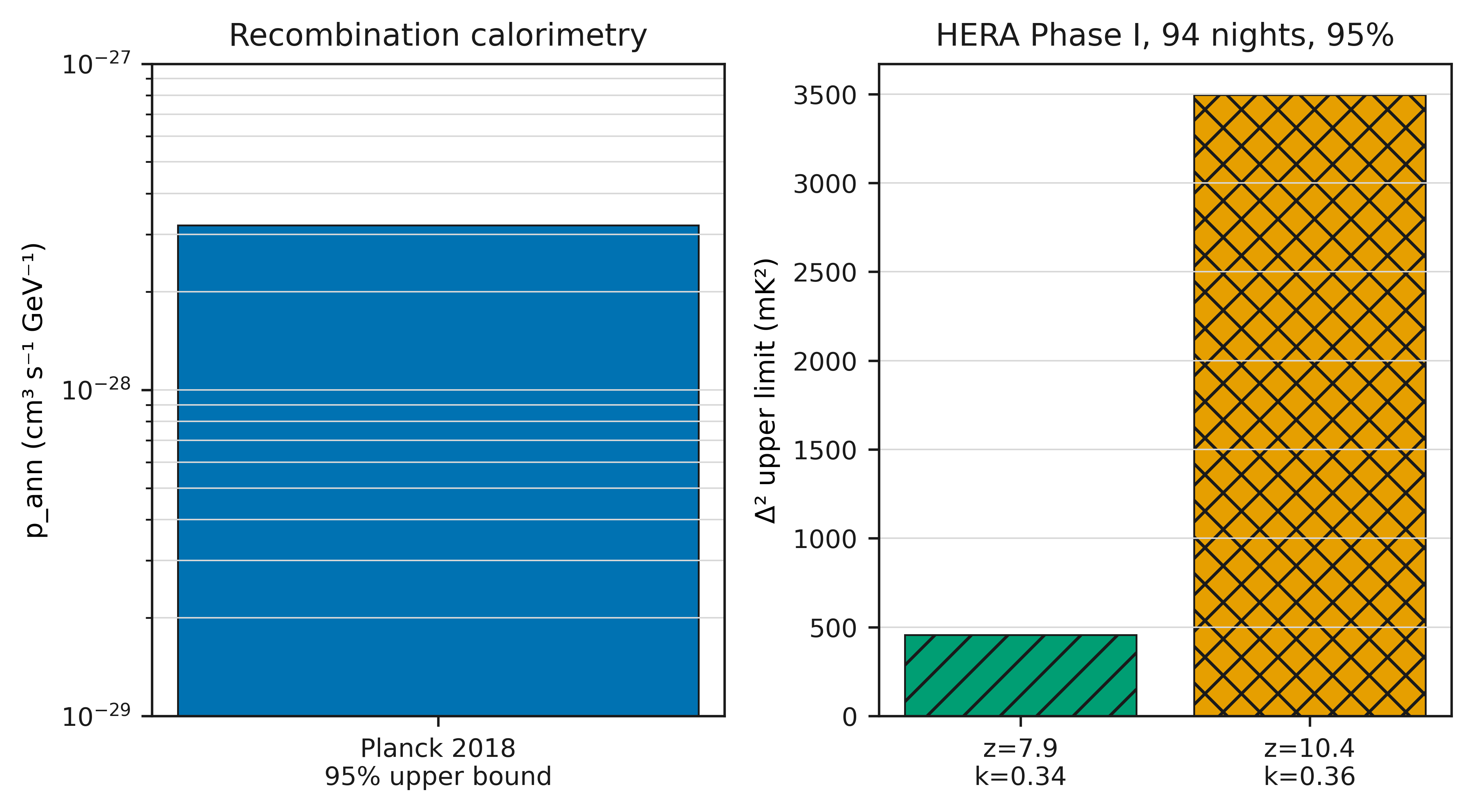}
\caption{Early-Universe quantitative constraints. The Planck \(p_{\mathrm{ann}}\) bound is shown separately from HERA power-spectrum upper limits and model-conditioned 21-cm forecasts. Sources: \cite{ref6,ref154,ref153,ref156,ref158}.}
\label{fig:early-universe}
\end{figure}

Fig.~\ref{fig:early-universe} illustrates the complementarity of the two epochs: the CMB
probes energy deposition near recombination, whereas 21-cm observations
probe later thermal and ionization evolution. Their different redshift
sensitivities are particularly useful for velocity-dependent or delayed
energy injection.

\section{Systematics, Statistical Validation, and Reproducible Combinations}\label{systematics-statistical-validation-and-reproducible-combinations}

Indirect DM searches are often limited by nuisance modeling
rather than by counting statistics. Physical backgrounds, such as
unresolved source populations or secondary particle production, should
be constrained using auxiliary data, whereas instrumental and modeling
uncertainties require calibration or alternative forward models.
Treating these effects as a single normalization uncertainty can
underestimate correlated residuals. Recent flexible analyses illustrate
this point: a 2025 Gaussian-process study of the Galactic-center excess
found that its inferred morphology depends appreciably on the adopted
foreground model and the region of interest, while HERA Phase II
identified mutual coupling as a dominant systematic that can transfer
foreground power into cosmological modes \cite{ref152}.

Statistical calibration is equally important. Signal strengths are
bounded at zero, nuisance parameters may be weakly identified under the
null hypothesis, and scans over mass, target, morphology, or region of
interest produce look-elsewhere effects. Local significances should
therefore be accompanied by the effective global search scope, and
simulations should validate asymptotic approximations when counts are
low, or nuisance behavior is nonstandard. Signal-injection and coverage
tests should probe both alternative backgrounds and plausible signal
misspecification. Modern simulation-based inference frameworks
increasingly provide explicit posterior-coverage diagnostics, but their
validity remains conditional on the simulations spanning the relevant
data-generating processes \cite{ref159}.

Combinations should be constructed from likelihoods rather than from
overlaid exclusion curves. Shared quantities---including halo factors,
particle spectra, production cross sections, and common calibration
inputs---must be represented only once, while detector-specific
backgrounds remain independent. Reproducible analyses should therefore
provide likelihoods or sufficient statistical representations, response
information, nuisance definitions and correlations, and documented
validation tests. These shared nuisance parameters and statistical choices propagate across messenger classes. Table~\ref{tab:cross_cutting_statistical_controls} summarizes the cross-messenger stress tests and reporting requirements needed to assess statistical robustness.

\setlength{\LTcapwidth}{\columnwidth}
\setlength{\tabcolsep}{2pt}
\renewcommand{\arraystretch}{1.15}
\setlength{\emergencystretch}{2em}

\footnotesize

\begin{longtable}[]{@{}
  >{\raggedright\arraybackslash}p{(\columnwidth - 6\tabcolsep) * \real{0.19}}
  >{\raggedright\arraybackslash}p{(\columnwidth - 6\tabcolsep) * \real{0.33}}
  >{\raggedright\arraybackslash}p{(\columnwidth - 6\tabcolsep) * \real{0.26}}
  >{\raggedright\arraybackslash}p{(\columnwidth - 6\tabcolsep) * \real{0.22}}
@{}}

\caption{Cross-cutting statistical validation and robustness controls for indirect DM searches, highlighting key stress tests, minimum reporting requirements, and common failure modes across multiple analysis strategies. Sources: \cite{ref7,ref24,ref63,ref66,ref67,ref82,ref160,ref161}.}
\label{tab:cross_cutting_statistical_controls} \\

\toprule
\textbf{Analysis issue} &
\textbf{Required stress test} &
\textbf{Minimum reporting} &
\textbf{Failure exposed} \\
\midrule
\endfirsthead

\toprule
\textbf{Analysis issue} &
\textbf{Required stress test} &
\textbf{Minimum reporting} &
\textbf{Failure exposed} \\
\midrule
\endhead

\midrule
\multicolumn{4}{r}{\footnotesize Continued on next page} \\
\endfoot

\bottomrule
\endlastfoot


Background model &
Alternative physically motivated models and signal injection &
Signal shift and nuisance covariance &
Background absorbs or creates a signal \\

Instrument response &
Calibration/control samples and response variations &
Scale, resolution, and response nuisance parameters &
Instrumental feature mimics signal \\

Model flexibility &
Alternative signal and background families &
Sensitivity to model class &
Overfitting or signal absorption \\

Physical boundary / low counts &
Monte Carlo coverage under the null &
Confidence construction and coverage &
Invalid asymptotic significance \\

Trials &
Full scan over mass, ROI, targets, and templates &
Local and global significance &
Hidden look-elsewhere effect \\

Combination &
Shared-nuisance and overlapping-data audit &
Joint likelihood and correlations &
Double counting or inconsistent priors \\

Simulation / ML inference &
Coverage and out-of-distribution tests &
Simulator, priors, and calibration diagnostics &
Domain shift or miscalibrated posterior \\

\midrule

\multicolumn{4}{@{}p{\columnwidth}@{}}{%
\footnotesize
\textit{Note.} Messenger-specific backgrounds and detector systematics
are discussed in the corresponding channel sections; Table~\ref{tab:cross_cutting_statistical_controls}
summarizes only controls that apply across multiple search strategies.
} \\

\end{longtable}

\normalsize

\section{A Multi-Messenger Discovery Framework}\label{a-multi-messenger-discovery-framework}

A multi-messenger DM analysis should begin with a
common-particle hypothesis---mass, annihilation cross section or
lifetime, branching fractions, and relevant velocity or mediator
dependence---rather than with independent signal normalizations. Link
shared astrophysical quantities, such as halo parameters, across
targets, while keeping messenger-specific transport, foreground, and
detector nuisances separate.

The strongest combinations are not necessarily those that yield the
tightest numerical limits, but those in which independent data constrain
different nuisance directions. For example, dwarfs, the CMB, neutrinos,
and charged particles can test interpretations of a Galactic-center
signal without sharing its dominant diffuse-emission uncertainties. Null
results are therefore essential to discovery inference when they
constrain the unavoidable counterparts of a candidate model.

Cross-messenger combinations should use consistent particle spectra, halo assumptions, and correlated nuisance parameters. Where possible, analyses should preserve the underlying likelihoods, instrumental responses, and covariance; combining digitized exclusion curves or treating shared astrophysical uncertainties as independent can produce artificial precision. Prospective tests are especially valuable because they restrict a model's flexibility after an anomaly has been identified.

We propose five conditions for a credible indirect-detection discovery:

\begin{enumerate}
\def\labelenumi{\arabic{enumi}.}
\item
  Instrumental replication: the feature appears in an independent
  detector or analysis with different dominant systematics.
\item
  Target scaling: intensity and morphology follow the predicted
  J-factor, D-factor, capture rate, conversion probability, or
  cosmological kernel.
\item
  Spectral closure: lines, continua, endpoints, and secondary particles
  are consistent with one particle model.
\item
  Nuisance orthogonality: at least one confirming data set does not
  share the leading systematic of the original signal.
\item
  Prospective prediction: the combined model correctly predicts a signal
  or absence of signal in previously unexamined data.
\end{enumerate}

Current anomalies do not yet satisfy all five conditions. Their
principal value is therefore not as isolated discovery claims but as
drivers of improved background models, calibration, environmental
measurements, and independent tests. A convincing multi-messenger
discovery will emerge when a single-particle model survives these
increasingly restrictive cross-target and cross-instrument predictions.

\section{Future Experimental Landscape and Research Priorities}\label{future-experimental-landscape-and-research-priorities}

Future progress in indirect DM searches will depend as much on
reducing systematic uncertainties as on increasing exposure. CTAO will
extend Galactic-center and dwarf searches into the multi-TeV regime,
while SWGO will add wide-field southern-sky coverage and sensitivity to
extended halos \cite{ref162,ref163,ref164}. Current HAWC and LHAASO
measurements already demonstrate the reach of TeV--PeV diffuse emission
for very heavy DM. In the MeV band, COSI-class instruments
target sub-GeV DM, PBHs, and positron-related emission, where
activation backgrounds and extended-source reconstruction are critical
\cite{ref165,ref166}. IceCube Upgrade, IceCube-Gen2, and KM3NeT provide
complementary improvements in threshold, volume, and detector medium
\cite{ref102,ref103,ref124}, while GAPS adds an independent
low-energy antinucleus technique \cite{ref167}. XRISM has already
demonstrated the value of high-resolution calorimetry for keV-line tests
\cite{ref130}, and HERA Phase II and SKA-Low extend early-Universe searches
through increasingly resolved 21-cm information \cite{ref152,ref157}. Table~\ref{tab:priority_dm_facilities} summarizes the priority experimental programs alongside their most important discovery-enabling advances and systematic requirements.

\setlength{\LTcapwidth}{\columnwidth}
\setlength{\tabcolsep}{2pt}
\renewcommand{\arraystretch}{1.15}
\setlength{\emergencystretch}{2em}

\footnotesize

\begin{longtable}[]{@{}
  >{\raggedright\arraybackslash}p{(\columnwidth - 4\tabcolsep) * \real{0.19}}
  >{\raggedright\arraybackslash}p{(\columnwidth - 4\tabcolsep) * \real{0.36}}
  >{\raggedright\arraybackslash}p{(\columnwidth - 4\tabcolsep) * \real{0.45}}
@{}}

\caption{Priority facilities and discovery-enabling advances for indirect DM searches, highlighting their principal scientific contributions and the key systematic requirements that govern discovery sensitivity. Sources: \cite{ref10,ref25,ref37,ref51,ref58,ref59,ref103,ref104,ref106,ref107,ref113,ref119,ref156,ref157,ref172}.}
\label{tab:priority_dm_facilities} \\

\toprule
\textbf{Facility / program} &
\textbf{Principal DM contribution} &
\textbf{Critical systematic / requirement} \\
\midrule
\endfirsthead

\toprule
\textbf{Facility / program} &
\textbf{Principal DM contribution} &
\textbf{Critical systematic / requirement} \\
\midrule
\endhead

\midrule
\multicolumn{3}{r}{\footnotesize Continued on next page} \\
\endfoot

\bottomrule
\endlastfoot


CTAO &
Galactic-center continuum and line searches, and dwarf-galaxy searches
from \(\sim\)TeV to \(>100\)~TeV &
Diffuse foregrounds, energy calibration, and halo/\(J\)-factor
uncertainty \\

SWGO &
Wide-field searches of the southern Galactic halo, dwarf galaxies,
and very-heavy DM &
Large-scale acceptance and hadronic-background stability \\

HAWC / LHAASO &
Current TeV--PeV annihilation and decay constraints &
Prompt and secondary emission, absorption, and diffuse backgrounds \\

COSI / next-generation MeV missions &
Sub-GeV DM, PBHs, 511-keV emission, and MeV continua &
Activation lines, response tails, and extended-source reconstruction \\

IceCube Upgrade / Gen2 &
Solar capture, Galactic DM, and high-mass decay searches &
Ice calibration, energy/angular response, and astrophysical-neutrino
background \\

KM3NeT &
Independent Galactic-center and solar DM tests &
Water optical response, atmospheric backgrounds, and exposure growth \\

GAPS &
Low-energy antiproton and antinucleus searches &
Production/coalescence uncertainties and instrumental backgrounds \\

XRISM / future calorimeters &
Sterile-neutrino and narrow X-ray line searches &
Gain calibration, atomic-line modeling, and target scaling \\

HERA / SKA-Low &
Exotic heating and ionization during cosmic dawn &
Foreground leakage, instrumental coupling, and astrophysical priors \\

\end{longtable}

\normalsize

Across all channels, the highest-leverage infrastructure is likelihood
portability: public response matrices, covariance, halo posteriors,
particle spectra, and nuisance correlations would allow new candidate
models to be tested consistently without reconstructing results from
exclusion curves. Future experiments should therefore pair each
sensitivity gain with a quantitative reduction of the nuisance parameter
that currently limits interpretation.

\section{Conclusions}\label{conclusions}

Indirect DM detection has become a precision, multi-messenger
field, yet no signal has reached discovery-level significance. Gamma
rays provide some of the strongest annihilation limits; charged
particles offer precision spectral tests conditioned on transport;
neutrinos add directional information and a unique connection to solar
capture; X-ray and radio observations probe decay and secondary
emission; and CMB and 21-cm measurements constrain energy injection
across cosmic time. The breadth of viable candidates---including WIMPs,
PBHs, sterile neutrinos, ALPs, dark photons, secluded sectors, and
superheavy relics---requires interpretation beyond a single benchmark
channel or cross section.

The central conclusion is that sensitivity alone does not establish evidence for dark matter. A credible
signal must remain stable across alternative background and nuisance
models, account for trial and instrumental effects, follow the expected
target or spatial scaling, and yield consistent predictions on
independent data. Likewise, exclusion limits are most useful when their
dependence on halo models, propagation, detector response, and
statistical assumptions is made explicit. Machine-learning and
simulation-based methods can improve inference, but only when coverage,
calibration, and robustness to model misspecification are demonstrated.

Progress will depend on both improved sensitivity and more reproducible interpretation. Public likelihoods, response information, covariance,
shared particle spectra, and consistent treatment of correlated
astrophysical uncertainties will be increasingly important for combining
experiments. The most convincing discovery is unlikely to be defined by
the largest isolated excess or the lowest exclusion curve, but by a
particle-level hypothesis that yields successful, testable predictions
across targets, instruments, and messengers.

\section*{Acknowledgements}

This review was partially supported by the College of Natural
and Health Sciences at Zayed University through the Research
Incentive Fund (Grant No. 23048), awarded to Dr. Muad Ghaith.
The authors gratefully acknowledge this financial support.

\end{document}